\documentclass[aps,pra,twocolumn]{revtex4}
\usepackage{graphicx}

\newcommand{\brs}{{\bf r}}
\newcommand{\brc}{{\bf R}}
\newcommand{\br}{{\bf r}}
\newcommand{\ec}{\underline{\underline{\brs}}}
\newcommand{\nc}{\underline{\underline{\brc}}}
\newcommand{\ncs}{\underline{\underline{s}}}
\newcommand{\ecs}{\underline{\underline{\sigma}}}
\newcommand{\nceq}{\underline{\underline{\brc}}_{eq}}

\newcommand{\hxc}{{\rm Hxc}}
\newcommand{\ehc}{E_{{\rm Hc}}^{en}}
\newcommand{\uhxc}{{\rm U,Hxc}}
\newcommand{\hc}{{\rm Hc}}
\newcommand{\h}{\hat{H}}
\newcommand{\p}{\varphi}
\newcommand{\den}{[\Gamma_,\rho]}
\newcommand{\ehxc}{E_{\rm Hxc}^{en}}

\newcommand{\htp}{$H_2^+\:$}
\newcommand{\asneq}{\stackrel{ {\scriptstyle R \to \infty} } 
{\longrightarrow}}

\begin{document}

\title{Multicomponent Density-Functional Theory for Electrons and Nuclei}
\author{Thomas Kreibich}
\affiliation{Institut f\"ur Theoretische Physik, Universit\"at W\"urzburg,
  Am Hubland, D-97074 W\"urzburg, Germany}
\author{Robert van Leeuwen}
\affiliation{Theoretical Chemistry, Materials Science Centre, University of Groningen,
Nijenborgh 4, 9747AG, Groningen, The Netherlands}
\author{E.K.U.Gross}
\affiliation{Institut f\"{u}r Theoretische Physik, Freie Universit\"{a}t Berlin, 
Arnimallee 14, D-14195 Berlin, Germany}

\date{\today}
\begin{abstract}
We present a general multi-component density functional theory
in which electrons and nuclei are treated completely quantum mechanically,
without the use of a Born-Oppenheimer approximation. The two fundamental
quantities in terms of which our theory is formulated are the nuclear N-body density and
the electron density expressed in coordinates referring to the nuclear framework. 
For these two densities coupled
Kohn-Sham equations are derived and the electron-nuclear correlation
functional is analyzed in detail. The formalism is tested on the hydrogen molecule $H_2$
and its positive ion $H_2^+$ using several approximations
for the electron-nuclear correlation functional.

\end{abstract}

\maketitle

\section{Introduction}
\label{sec:intro}

Density functional theory (DFT) is among the most succesful approaches 
to calculate the electronic structure of atoms, molecules and solids.
In its original form~\cite{HohenbergKohn,KohnSham}, DFT always invokes the
Born-Oppenheimer approximation: One is supposed to calculate the electron
density $\rho ( \brs)$ which is in 1-1 correspondence to the static
potential potential of {\em fixed} nuclei.
In a recent Letter~\cite{KreibichGross:2001} we introduced 
a multicomponent density-functional theory (MCDFT) for the complete 
quantum treatment
of many-particle systems consisting of electrons and nuclei.
With this theory it is possible to decribe from first principles 
physical phenomena that depend on a strong coupling between electronic and nuclear
motion. MCDFT thereby extends the widely applied density 
functional formalism for purely electronic properties,
opening up a new field of applications,
such as the first-principles calculation of electron-phonon
coupling in solids~\cite{vanLeeuwen:PRB04} which is a key ingredient in the
description of superconductivity~\cite{supercond1,supercond2,supercond3,supercond4} and 
polaronic motion~\cite{Mahan:book,HannewaldBobbert:PRB04}.
The quantum treatment of the nuclear motion in molecules or solids 
is essential in situations that from a Born-Oppenheimer (BO)
viewpoint must be described by a 
superposition of different BO structures. 
This is, for instance, 
the case in floppy molecules~\cite{BacicLight:ARPC89}, or in so-called switchable
molecules\cite{Dulicetal:PRL03} which are in a superposition of an open and a closed
state after a laser excitation. Apart from treating such various
phenomena the MCDFT presented here also paves the way
for future time-dependent extensions of the theory
which would enable one to calculate the
coupled electronic and nuclear dynamics of many-particle
systems, within linear response and beyond. 
Indeed, some preliminary steps towards the description
of the coupled ionization and dissociation dynamics of
molecules in strong laser fields have already been 
taken~\cite{Kreibich:thesis,Nikitasvolume,Kreibichetal:CP04}.\\
The purpose of the present work is twofold. First, we want to give an
extended and detailed description of the theory that was briefly
described in our Letter. Second, we want to investigate
in detail some new approximate density functionals for the electron-nuclear
correlation and see how they perform. To do this the formalism is
tested on the hydrogen molecule and its positive ion.
The paper is organized as follows. In section II we first introduce the
basic formalism and discuss the Hohenberg-Kohn theorem and
the Kohn-Sham equations in a multi-component theory in which the 
electron density is defined with respect to
a coordinate frame attached to the nuclear framework and in
which the diagonal of the nuclear density matrix appears as a new variable. 
In section III we perform an analysis of the several energy
functionals and of the resulting potentials in the Kohn-Sham equations.
Furthermore, the connections between the effective potential of the nuclear 
Kohn-Sham equation and the Born-Oppenheimer energy surface is analyzed.
In section IV we apply our formalism and test several approximate
forms for the electron-nuclear correlation functional for
the case of the hydrogen molecule and its positive ion.
Finally in section V we present our conclusions.

\section{Basic Formalism}
\label{sec:formalism}
\subsection{Discussion of the Hamiltonian}
\label{ssec:ham}

We consider a system composed of $N_e$ electrons with coordinates 
$\{\brs_j\} \equiv \ec$ and $N_n$ nuclei with masses $M_1 ... M_{N_n}$, 
charges $Z_1 ... Z_{N_n}$, and coordinates denoted by $\{\brc_{\alpha}\} 
\equiv \nc$.
By convention, the subscript ``e'' and ``n'' refer to electrons and nuclei, 
respectively, and atomic units are employed throughout this work.
In non-relativistic quantum mechanics, the system is 
decribed by the Hamiltonian
\begin{eqnarray}
  \label{eq:ham-inert}
  \h &=& \hat{T}_{\rm n}(\nc) + \hat{W}_{nn}(\nc)
  + \hat{U}_{{\rm ext},n}(\nc) \nonumber \\
  &+& \; \hat{T}_{e}(\ec) \;\, + \, \hat{W}_{ee}(\ec)
  \;\, + \, \hat{U}_{{\rm ext},e}(\ec) \nonumber \\
  &+&  \hat{W}_{en}(\nc,\ec),
\end{eqnarray}
where
\begin{eqnarray}
  \label{eq:ekin}
  \hat{T}_{n} &=& \sum_{\alpha=1}^{N_n} \left( -\frac{\nabla^2_{\alpha}}
    {2 M_{\alpha}} \right) \\
  \hat{T}_{e} &=& \sum_{j=1}^{N_e} \left( -\frac{\nabla^2_j}{2} \right)
\end{eqnarray}
denote the kinetic-energy operators of the nuclei and
electrons, respectively, and
\begin{eqnarray}
  \label{eq:w1}
  \hat{W}_{nn} &=& \frac{1}{2} \sum_{\stackrel{\alpha,\beta=1}
    {\alpha \not= \beta}}^{N_n} \frac{Z_{\alpha}Z_{\beta}}
  {|\brc_{\alpha}-\brc_{\beta}|} \\
  \hat{W}_{ee} &=& \frac{1}{2} \sum_{\stackrel{i,j=1}
    {i \not= j}}^{N_e} \frac{1}{|\br_i-\br_j|} \\
  \label{eq:w}
  \hat{W}_{en} &=& - \sum_{j=1}^{N_e} \sum_{\alpha=1}^{N_n}
  \frac{Z_{\alpha}}{|\br_j - \brc_{\alpha}|}
\end{eqnarray}
represent the interparticle Coulomb interactions.
We emphasize that no BO approximation has been assumed in~(\ref{eq:ham-inert});
the Hamiltonian of Eq.(\ref{eq:ham-inert}) provides a quantum mechanical
description of all, i.e., electronic and nuclear degrees of freedom.
In contrast to the standard approach using the BO approximation,
the interactions between electrons and nuclei are therefore
treated within $\hat{W}_{en}$, Eq.~(\ref{eq:w}),
and do not contribute to the external potentials.
Truly external potentials representing, e.g., a 
voltage applied to the system, are contained in
\begin{eqnarray}
  \label{eq:uext}
  \hat{U}_{{\rm ext},n} &=& \sum_{\alpha=1}^{N_n} U_{{\rm 
ext},n}(\brc_{\alpha}) \\
   \hat{U}_{{\rm ext},e} &=& \sum_{j=1}^{N_e} u_{{\rm ext},e}(\br_j).
\end{eqnarray}
Defining electronic and nuclear single-particle densities 
conjugated to the true external potential (\ref{eq:uext}),
a MCDFT formalism can readily be formulated on the basis of the 
above Hamiltonian \cite{CapitaniNalewajskiParr:82}.
However, as discussed in \cite{KreibichGross:2001}, such a MCDFT is not 
useful in practice because the single-particle densities necessarily 
reflect the symmetry of the true external potentials and are therefore not
characteristic of the internal properties of the system.
In particular, for all isolated systems where the external potentials
(\ref{eq:uext}) vanish, these densities, as a consequence of the translational
 invariance of the respective Hamiltonian, are constant.

A suitable MCDFT is obtained by defining the densities with respect to internal
coordinates of the system \cite{KreibichGross:2001}.
To this end, new electronic coordinates are introduced according to
\begin{equation}
  \label{eq:rep}
  \br'_j = \mathcal{R}(\alpha,\beta,\gamma) \; \left( \br_j - \brc_{\rm CMN}
	\right) \qquad j = 1...N_e,
\end{equation}
where
\begin{equation}
  \label{eq:rcmn}
   \brc_{\rm CMN} := \frac{1}{M_{\rm nuc}} \sum_{\alpha=1}^{N_n}
    M_{\alpha} \brc_{\alpha} .
\end{equation}
denotes the center of mass (CM) of the nuclei, the total nuclear mass is
given by
\begin{equation}
  \label{eq:mnuc}
  M_{\rm nuc} = \sum_{\alpha=1}^{N_n} M_{\alpha},
\end{equation}
and $\mathcal{R}$ is the three-dimensional orthogonal matrix representing
the Euler rotations \cite{Goldstein:80}.
The Euler angles $(\alpha,\beta,\gamma)$ are functions of the
nuclear coordinates $\{ \nc \}$ and specify 
the orientation of the body-fixed coordinate frame.
They can be determined in various ways. One way to define them is by requiring the
inertial tensor of the nuclei to be diagonal in the body-fixed 
frame. The conditions that the off-diagonal elements of the inertia tensor are
zero in terms of the rotated coordinates
$\mathcal{R}(\brc_{\alpha}-\brc_{CMN})$ 
then give three determining equations for the three Euler angles in terms
of the nuclear coordinates $\{ \nc \}$.
This way of choosing the Euler angles is commonly used within the field of
nuclear physics~\cite{Villars:57,VillarsCooper:70,Bucketal:79} but is, of course,
not unique. 
A common alternative way to determine the orientation
of the body-fixed system is provided by the so-called Eckart conditions
\cite{Eckart:35,Wilsonetal:55,LouckGalbraith:76,Louck:76,Bunker:79,Sutcliffe:93,Sutcliffe:00} which are suitable to
describe small vibrations in molecules and phonons in solids \cite{vanLeeuwen:PRB04}.  
A general and very elegant discussion on the various ways the
body-fixed frame can be chosen is given in reference \cite{LittlejohnReinsch:97} .
In this work we will not make a specific choice as our derivations are independent 
of such a choice. The most important point is that
by virtue of Eq.~(\ref{eq:rep}), the electronic coordinates are 
defined with respect to a coordinate frame that is attached to the 
nuclear framework and rotates as the nuclear framework rotates.
In fact, this transformation comprises two transformations:
A first one transforming the {\it space-fixed inertial} coordinates into 
{\it CM-fixed relative} coordinates, and a second one transforming the
{\it CM-fixed relative} coordinates into {\it body-fixed internal} 
coordinates.

The nuclear coordinates themselves are not transformed any further
at this point, i.e.,
\begin{equation}
  \label{eq:rndp}
  \brc_{\alpha}' = \brc_{\alpha} \qquad \alpha=1...N_n  .
\end{equation}
Of course, introducing internal nuclear coordinates is also desirable.
However, the choice of  such coordinates depends strongly on
the specific system to be described:
If near-equilibrium situations in systems with well-defined geometries
are considered, normal or -- for a solid --
phonon coordinates are most appropriate, whereas fragmentation processes
of molecules are better described in terms of Jacobi coordinates
\cite{Schinke:93}.
Therefore, keeping a high degree of flexibility,
the nuclear coordinates are left unchanged for the time being and are
only transformed to internal coordinates  prior to actual applications in the final
equations that we will derive.
Another reason for not introducing any internal nuclear coordinates
at this point, is to retain simple forms of the equations.
In a transformation to internal nuclear coordinates typically the nuclear
center-of-mass and the Euler angles are taken as new variables
as well as $3N_n-6$ internal or shape 
coordinates $Q_i$ \cite{LittlejohnReinsch:97,Sutcliffe:93,Sutcliffe:00}. 
These internal coordinates, however, do not have a simple relation to the
original $N_n$ nuclear coordinates and will therefore lead to a complicated
form of the Hamiltonian in the new coordinates. We will therefore delay the use 
of such transformations until we have derived the final equations.

As a result of the coordinate changes of Eq.(\ref{eq:rep}) , the Hamiltonian
(\ref{eq:ham-inert}) transforms into
\begin{eqnarray}
  \label{eq:ham-trafo}
  \h &=& \hat{T}_{n}(\nc) + \hat{W}_{nn}(\nc) 
	+ \hat{U}_{{\rm ext},n}(\nc)  \nonumber \\
     &+& \hat{T}_{e}(\ec') + \hat{W}_{ee}(\ec') + \hat{T}_{\rm MPC}(\nc,\ec')\nonumber \\
  &+& \hat{W}_{en}(\nc,\ec')
	+ \hat{U}_{{\rm ext},e}(\nc,\ec') .
\end{eqnarray}
Since we have transformed to a noninertial 
coordinate frame mass-polarization and Coriolis (MPC) terms
\begin{equation}
  \hat{T}_{\rm MPC} := \sum_{\alpha=1}^{N_n} 
	-\frac{1}{2 M_{\alpha}} 
  \left( \nabla_{\brc_\alpha} + \sum_{j=1}^{N_e} \frac{ \partial \br_j'}
    {\partial \brc_{\alpha}} \nabla_{\br_j'} \right)^2 - \hat{T}_n(\nc) 
 \label{eq:tmpcdef1}
\end{equation}
appear.
Obviously, $\hat{T}_{\rm MPC}$ is not symmetric in the electronic and nuclear 
coordinates.
However, this was not expected since only the electrons refer to a 
noninertial coordinate frame, whereas the nuclei are still defined with 
respect to the inertial frame.
Therefore, all MPC terms arise solely from the electronic coordinates,
representing ficticious forces due to the electronic motion in noninertial 
systems (for a detailed form of these terms within the current coordinate
transformation see~\cite{vanLeeuwen:PRB04}) .
The kinetic-energy operators $\hat{T}_e$ and $\hat{T}_n$,
the electron-electron and nuclear-nuclear interactions, as well as the
true external potential $\hat{U}_{{\rm ext},n}$ acting on the nuclei 
are formally unchanged in Eq.~(\ref{eq:ham-trafo}) and therefore given by
Eqs.~(\ref{eq:ekin}) and (\ref{eq:w1}) with the new coordinates
replacing the old ones, whereas the electron-nuclear
interaction now reads
\begin{eqnarray} 
\lefteqn{\hat{W}_{en}(\nc,\ec') }
\nonumber \\
&=& - \sum_{j=1}^{N_e} \sum_{\alpha=1}^{N_n}
  \frac{Z_{\alpha}}
{|\mathcal{R}(\alpha,\beta,\gamma)^{-1}\br_j'-\brc_{\alpha}+\brc_{CMN}|}  \nonumber \\
    &=& - \sum_{j=1}^{N_e} \sum_{\alpha=1}^{N_n}
  \frac{Z_{\alpha}}{| \br_j'- \mathcal{R} (\alpha,\beta,\gamma) (\brc_{\alpha}- \brc_{CMN})|} 
 \label{eq:wen2def}.
\end{eqnarray}
The quantity 
\begin{equation}
\brc_{\alpha}'' = \mathcal{R}(\alpha,\beta,\gamma) (\brc_{\alpha} - \brc_{CMN})
\end{equation}
that appears in Eq.(\ref{eq:wen2def})
is a so-called shape coordinate~\cite{LittlejohnReinsch:97,vanLeeuwen:PRB04}, i.e.
it is invariant under rotations and translations of the nuclear framework
\begin{equation}
\brc_{\alpha}'' (O \nc + \mathbf{a} ) = \brc_{\alpha}'' ( \nc)
\label{eq:shape}
\end{equation}
where $O$ is an arbitrary rotation matrix and $\mathbf{a}$ an arbitrary
translation vector. The invariance property described in Eq.(\ref{eq:shape})
is simply a consequence of the fact that the Euler angles are
defined by giving the vectors $\brc_{\alpha}''$ certain values,
independent of where the nuclear center-of-mass was situated in the
laboratory frame or how the nuclear framework was orientated.
This is, of course, precisely the purpose of introducing a
body-fixed frame. For this reason the potential in Eq.(\ref{eq:wen2def}) that
the electrons in the body-fixed frame experience from the nuclei is invariant
under rotations or translations of the nuclear framework.

As a further result of the coordinate transformation (\ref{eq:rep}),
the true external potential acting on the electrons now not only depends on 
the electronic coordinates, but also on all the nuclear coordinates:
\begin{equation}
  \label{eq:uext-trafo}
  \hat{U}_{{\rm ext},e}(\nc,\ec') = \sum_{j=1}^{N_e} 
	u_{{\rm ext},e}(\mathcal{R}^{-1}\br_j' +\brc_{CMN}).
\end{equation}
In the chosen coordinate system the electron-nuclear interaction~(\ref{eq:wen2def})
and the external potential~(\ref{eq:uext-trafo}) remain one-body
operators with respect to the electronic degrees of freedom but
represent complicated $N_n$-body interactions with respect to the nuclei.
We finally discuss some general aspects of our coordinate transformation.
If we consider the symmetry properties of our original Hamiltonian
of Eq.(\ref{eq:ham-inert}) in the absence of external potentials, we see that
it is invariant under simultaneous translations and rotations of all particles,
i.e. of both electrons and nuclei. This is not true anymore for our transformed
Hamiltonian.
Since we transformed the electronic coordinates to a body-fixed frame
we find that in the absence of external potentials the transformed
Hamiltonian of Eq.(\ref{eq:ham-trafo}) is invariant under translations
and rotations of nuclear coordinates only. The corresponding ground-state
wavefunction, if it is nondegenerate, will have the same invariance.\\
Let us next consider the permutational symmetry.
The ground state wavefunction of the original Hamiltonian of Eq.(\ref{eq:ham-inert})
is antisymmetric under the interchange
of electronic space-spin coordinates and symmetric or antisymmetric
under interchange of nuclear space-spin coordinates of nuclei
of the same type, depending on whether they are bosons or fermions.
The ground state wavefunction of the transformed Hamiltonian of Eq.(\ref{eq:ham-trafo}) 
will also be antisymmetric with respect to the interchange of electronic 
space-spin coordinates. However, the symmetry properties with respect to
the interchange of the nuclear space-spin coordinates depend
on the conditions that we choose to determine the Euler angles.
If we choose a determining constraint for the Euler angles 
that is symmetric in the interchange of
particles of the same type, then the transformed wavefunction will retain
the permutational symmetry properties of the original wavefunction.
This is, for instance, the case if we determine the Euler angles
by the requirement that the nuclear inertia tensor be diagonal.
However, if we choose a nonsymmetric constraint, such as the
Eckart conditions, then the transformed wavefunction will have more
complicated transformation properties under the interchange
of nuclear spin-space coordinates since the interchange of two
nuclear coordinates will then also change the Euler angles
(a detailed account on this topic is given in 
Ref.\cite{Sutcliffe:00}).
This can lead to practical complications
but will not affect our general formalism.\\
We finally note that the coordinate transformation we presented here did not aim at
a separation of the constants of motion of the system (even for the case 
of isolated systems).
In contrast, the transformation (\ref{eq:rep}) was chosen such that
the new electronic coordinates reflect the internal symmetry of the system.
We thus arrive at a Hamiltonian which naturally lends
itself as a starting point for the formulation of a MCDFT, as will be shown
in the subsequent sections.

\subsection{Definition of the Densities}
\label{ssec:dendef}

As a first step towards the formulation of a density functional theory,
one has to define the densities which will serve as the fundamental
variables of the theory.
Although this seems to be rather straightforward and is normally not
discussed at length, a careful definition of the densities
is of crucial importance in the current context.

As already mentioned above, it is not useful to define electronic and 
nuclear single-particle densities in terms of the inertial coordinates
$\br$ and $\brc$, since such densities necessarily reflect the symmetry of
the corresponding true external potentials, e.g., Galilean symmetry for 
vanishing external potentials.
Therefore, such single-particle densities are {\em not characteristic} for the internal
properties of the system under consideration. 

We proceed with the definition of a suitable set of densities, which
should fulfill the following requirements:
\begin{itemize}
\item
  They should be characteristic for the internal properties of the system;
  in particular, they should be meaningful in the limit of vanishing
  external potentials.
\item
  The basic electronic variable should be a single-particle quantity.
\item
  The treatment of the nuclear degrees of freedom should allow
  for appropriate descriptions of situations as different as
  near-equilibrium properties of solids and fragmentation processes of
  molecules.
\end{itemize}
A set of densities which meets these requirements is given by
\begin{eqnarray}
  \label{eq:gamman}
  \Gamma(\nc) &=& \hspace{1.5em} \sum_{s,\sigma} \int d^{N_e}\br' \;
  \left| \Psi(\nc \ncs,\ec' \ecs) \right|^2 \\
  \label{eq:dene}
  \rho(\br') &=& N_e \sum_{s,\sigma} \int d^{N_n}\brc \int d^{N_e-1}\br' \;
  \left| \Psi(\nc \ncs,\ec' \ecs) \right|^2
\end{eqnarray}
where $\Psi(\nc \ncs,\ec' \ecs)$ corresponds to the ground state of 
Hamiltonian (\ref{eq:ham-trafo}) and where $\ncs$ and $\ecs$ denote the
nuclear and electronic spin coordinates.
These densities are defined with respect to the transformed
coordinates $\{\nc,\ec'\}$.
In particular, the electronic single-particle density $\rho(\br')$
refers to the body-fixed molecular frame.
In terms these coordinates, the
quantity (\ref{eq:dene}) represents a conditional density, which is
characteristic for the internal properties of the system.
It is proportional to the probability density of finding an electron at
position $\br'$ as measured from the nuclear center-of-mass, given a
certain orientation of the nuclear framework.
Therefore the electronic density calculated through (\ref{eq:dene})
reflects the internal symmetries of the system, e.g., the cylindrical
symmetry of a diatomic molecule, instead of the Galilean symmetry
of the underlying space.
The nuclear degrees of freedom, on the other hand, 
are described using the diagonal of the nuclear density matrix, 
Eq.~(\ref{eq:gamman}). In the absence of external potentials
this quantity will have the transformation property
\begin{equation}
\label{eq:gam-trafo}
\Gamma ( O \nc + \mathbf{a}) = \Gamma (\nc)
\end{equation}
where $O$ is a rotation and $\mathbf{a}$ a translation vector.
Its permutational properties will depend on the choice of the
body fixed frame as discussed in the previous section.
The quantity $\Gamma (\nc)$ allows us to set up a general as well as flexible
formalism, which will be applicable to a large variety of situations.
In an actual application, one may at a later stage further contract this quantity
to obtain reduced density matrices or, depending
on the physical situation, introduce more suitable
internal nuclear coordinates which could not be done 
if single-particle
quantities had already been introduced at this point.

\subsection{The Hohenberg-Kohn Theorem for Multicomponent Systems}
\label{ssec:mchk}

In this section, we discuss the extension of the Hohenberg-Kohn theorem
to multicomponent systems.
In contrast to prior formulations of the MCDFT \cite{SanderShoreSham:73,%
KaliaVashishta:78,CapitaniNalewajskiParr:82,Gidopoulos:98,Ludena},
this analysis will employ the densities (\ref{eq:gamman}) and
(\ref{eq:dene}) as fundamental variables.
Correspondingly, the starting point of the following analysis is the
Hamiltonian (\ref{eq:ham-trafo}).
In order to formulate a Hohenberg-Kohn-(HK)-type statement, the  
Hamiltonian (\ref{eq:ham-trafo}) is generalized to
\begin{equation}
  \label{eq:ham}
  \h = \hat{T} + \hat{W} + \hat{U} + \hat{V} ,
\end{equation}
where
\begin{equation}
  \label{eq:ekintot}
  \hat{T}  = \hat{T}_{n}(\nc) + \hat{T}_{e}(\ec') 
	+ \hat{T}_{\rm MPC}(\nc,\ec')
\end{equation}
denotes the total kinetic-energy operator and
\begin{equation}
  \label{eq:wtot}
  \hat{W} = \hat{W}_{ee}(\ec')
  +  \hat{W}_{en}(\nc,\ec')
\end{equation}
contains the electron-electron and the electron-nuclear interaction.
Furthermore, auxiliary 'external' potentials conjugated to
the densities (\ref{eq:gamman}) and (\ref{eq:dene}),
\begin{equation}
  \label{eq:vtot}
  \hat{V} = \hat{V}_{n}(\nc) + \hat{V}_{e}(\ec'),
\end{equation}
have been added to the Hamiltonian.
We note that, in the transformed coordinates, $\hat{V}_{n}$ actually 
acts as an $N_n$-body operator with respect to the nuclear coordinates,
\begin{equation}
  \label{eq:potn}
  \hat{V}_{n} = V_n(\nc),
\end{equation}
and particularly contains the internuclear repulsion $\hat{W}_{nn} (\nc )$,
while $\hat{V}_{e}$ is a one-body operator with respect to the
(body-fixed) electronic coordinates:
\begin{equation}
  \label{eq:pote}
  \hat{V}_{e} = \sum_{j=1}^{N_e} v_e(\br'_j) .
\end{equation}
The 'true' external potentials, on the other hand, are subsumed in 
\begin{equation}
  \label{eq:utot}
  \hat{U} = \hat{U}_{{\rm ext},n}(\nc)
  +  \hat{U}_{{\rm ext},e}(\nc,\ec') .
\end{equation}
Note that the nuclear potential $\hat{U}_{{\rm ext},n}$ has the same 
structure as $\hat{V}_n$, whilst the electronic potential 
$\hat{U}_{{\rm ext},e}$ acts similar to the electron-nuclear interaction
in the transformed coordinate system.

The Hamiltonian (\ref{eq:ham}) and the above defined densities 
(\ref{eq:gamman}) and (\ref{eq:dene}) now provide a suitable basis for the
formulation of the multicomponent Hohenberg-Kohn (MCHK) theorem.
It can be summarized by the following statements:
\begin{enumerate}
\item
  {\it Uniqueness}:\\
  The set of ground-state densities $\{\Gamma,\rho\}$ uniquely determines
  the ground-state wavefunction, $\Psi = \Psi\den$, as well as 
   the potentials, $\{\hat{V}_{n} = \hat{V}_{n}\den,  
  \hat{V}_{e} = \hat{V}_{e}\den\}$.
  As a consequence, any observable of the static many-body system is
  a functional of the set of ground-state densities $\{\Gamma,\rho\}$.
\item
  {\it MCHK variational principle}:\\
  The total-energy functional
  \begin{equation}
    \label{eq:etot}
    E\den := \langle \Psi\den | \h | \Psi\den \rangle
  \end{equation}
  is equal to the exact ground-state energy $E_0$ if the exact densities
  $\Gamma_0$ and $\rho_0$ corresponding to fixed external potentials 
  $\hat{V}_{n,0}$ and $\hat{V}_{e,0}$ are inserted 
  into the functional.
  For all other densities, the inequality
  \begin{equation}
    \label{eq:mchkvar}
    E_0 < E \den
  \end{equation}
  holds true.
\end{enumerate}

This MCHK theorem can be proven by using both the {\it reductio ad 
absurdum}
and the {\it constrained search} approach, familiar from standard DFT
\cite{DreizlerGross:90}.
In the following, a generalization of the latter to
multi-component (MC) systems
will be presented.
We start out by defining the functional:
\begin{equation}
  \label{eq:fdef2}
  F\den := \min_{\Psi \to \Gamma,\rho} \langle
  \Psi | \hat{T} + \hat{W} + \hat{U} | \Psi \rangle ,
\end{equation}
i.e., we search for the minimum of $\langle \Psi | \hat{T} + \hat{W} + \hat{U} | \Psi
\rangle$ using all (properly normalized and symmetrized) wave functions
yielding a given set of densities $\{\Gamma,\rho\}$. It must be noted that all the
wave functions that we use in the constrained search procedure are now also required 
to have the correct symmetry properties respect to interchange
of nuclear space-spin coordinates of nuclei of the same type. As we discussed before
these symmetry properties depend on the way we define the body-fixed frame.
For instance, if we define the body-fixed frame by a diagonalization of the
nuclear inertia tensor then the constrained search must be carried out
over all wavefunctions that are antisymmetric in the electronic spin-space coordinates and
symmetric or anti-symmetric with respect to the interchange of
nuclear spin-space coordinates, depending on whether the nuclei are bosons
or fermions.
If we denote the minimizing state (assuming it exists~\footnote{In
standard electronic DFT, one can prove that the minimum
of $F$ exists \cite{Lieb:82}.})
by $\Psi^{\rm min}[\Gamma,\rho]$, we realize that
\begin{equation}
  \label{eq:fdef3}
  F[\Gamma,\rho] = \langle \Psi^{\rm min}\den | \hat{T} + \hat{W} + \hat{U} |
  \Psi^{\rm min}\den \rangle
\end{equation}
is -- by construction -- a functional of the densities.
We note that, in contrast to usual DFT, the functional $F$ is not universal since it
still depends on the external potentials $\hat{U}$ which, as a result of
our coordinate transformation, are functions of both $\nc$ and $\ec'$ 
as was discussed in connection with Eq.(\ref{eq:uext-trafo}).

Using Eq.(\ref{eq:fdef3}), the total-energy functional is given by
\begin{equation}
  \label{eq:etot2}
  E\den = F\den + \int d^{N_n}\brc \; \Gamma(\nc)
  V_n(\nc) + \int d\br \; \rho(\br) v_e(\br).
\end{equation}
The variational principle (\ref{eq:mchkvar}) can now be proven by 
employing the Rayleigh-Ritz variational principle:
\begin{equation}
  \label{eq:rrvar}
  E_0 = \min_{\Psi} \langle \Psi | \hat{H} | \Psi \rangle \:.
\end{equation}
Following the constrained-search procedure \cite{Levy:79}
of ordinary DFT, the minimum in (\ref{eq:rrvar}) is split
into two consecutive steps
\begin{eqnarray}
  \label{eq:etot3}
  E_0 &=& \min_{\Gamma,\rho} \left( \min_{\Psi \to \Gamma,\rho}
    \langle \Psi | \hat{H} | \Psi \rangle  \right)  \nonumber \\
    &=& \min_{\Gamma,\rho} \left( F\den + \int d^{N_n}\brc \;
      \Gamma(\nc)  V_n(\nc) \right. \nonumber \\
    && + \left. \int d\br \; \rho(\br) v_e(\br) \right) \nonumber \\
    &=& \min_{\Gamma,\rho} E\den,
\end{eqnarray}
where the external potentials $V_n$ and $v_e$
are held fixed during the minimization
(For notational simplicity, the primes indicating the transformed 
coordinates are dropped from now on.
By convention, all electronic coordinates are understood to refer to the 
body-fixed frame).
In the second step, we have exploited the fact that all wave functions
which lead to the same densities also yield the same external energy.
By virtue of the Rayleigh-Ritz variational principle, the minimizing
densities are the ground-state densities $\Gamma_0$ and $\rho_0$.
Furthermore, any other set of densities will lead to an energy above the
true ground-state energy if inserted in the total-energy functional
(\ref{eq:etot2}).
This completes the proof of statement 2.

In order to prove the first statement, we reformulate the variational
principle (\ref{eq:etot3}) according to
\begin{equation}
  \label{eq:etot4}
  \delta \Big\{ F\den + \int d^{N_n}\brc \; \Gamma(\nc)  V_n(\nc)
  + \int d\br \; \rho(\br) v_e(\br) \Big\} = 0 .
\end{equation}
Since the variations can be done independently, 
Eq.~(\ref{eq:etot4})
is equivalent to
\begin{eqnarray}
  \label{eq:etot5}
  \frac{\delta F\den }{\delta \Gamma(\nc)}  +
    V_n(\nc) &=& 0  \\
  \label{eq:etot6}
  \frac{\delta F\den }{\delta \rho(\br)}  + v_e(\br) &=& 0 \:.
\end{eqnarray}
If the exact densities $\{\Gamma_0,\rho_0\}$ are inserted,
the Euler equations (\ref{eq:etot5}) and (\ref{eq:etot6}) are satisfied
for the true external potentials.
If, on the other hand, an arbitrary set of densities $\{\Gamma,\rho\}$
is inserted,  Eqs.~(\ref{eq:etot5}) and (\ref{eq:etot6}) define -- assuming
the functional derivatives exist -- a set of
potentials, which reproduce $\{\Gamma,\rho\}$ as ground-state densities.
Therefore, the set of densities $\{\Gamma,\rho\}$ uniquely determines
the external potentials $\{V_n,v_e\}$ and thus the ground-state 
wavefunction $\Psi=\Psi^{\rm min}\den$.

Before concluding, a number of remarks are added:
\begin{itemize}
\item
  As usual, the potentials are uniquely determined up to an arbitrary
  additive constant, and non-degeneracy of the ground state has been 
  assumed.
\item
  Similar to purely electronic DFT, the functional $F\den$ is defined via 
  Eq.~(\ref{eq:fdef3}) for all
  {\it $\{\Gamma,\rho\}$-representable} densities, i.e. for all densities
  obtained according to Eqs.(\ref{eq:gamman}) and (\ref{eq:dene}) 
  from a many-body wave function with the right permutational symmetries.
  The potentials $\{V_n,v_e\}$ are defined for
  all densities, for which the functional derivatives in Eqs.~(\ref{eq:etot5})
  and (\ref{eq:etot6}) exist, i.e., for all {\it
    interacting $\{V_n,v_e\}$-representable} densities.
\item
  If vanishing external potentials (\ref{eq:uext-trafo}) are considered, 
  the analysis reduces to the one given in \cite{KreibichGross:2001}.
\end{itemize}

\subsection{The Kohn-Sham Scheme for Multicomponent Systems}
\label{ssec:mcks}

As usual, the HK theorem does not depend on the specific form of the
particle-particle interaction.
In particular, it can be applied to an auxiliary system which is
characterized by $\hat{W}=0$, i.e., the system consists of noninteracting
electrons and of nuclei that only interact amongst themselves.
The key assumption in establishing the MCKS scheme is
that local effective potentials
$\{\hat{V}_{S,n},\hat{V}_{S,e}\}$ exist such that the ground-state
densities of the auxiliary system reproduce the {\it exact}
ground-state densities $\{\Gamma_0,\rho_0\}$ of the fully interacting
system.
If that assumption holds true, the exact ground-state densities
are given by
\begin{eqnarray}
  \label{eq:ksnden}
  \Gamma_0(\nc) &=& \sum_{s} | \chi(\nc \ncs ) |^2 \\
  \label{eq:kseden}
  \rho_0(\br) &=& \sum_{j=1}^{N_e} | \p_j(\br) |^2 ,
\end{eqnarray}
where $\chi$ and $\p_j$ are solutions of an $N_n$-particle
nuclear and a single-particle electronic Schr\"odinger equation, 
respectively:
\begin{eqnarray}
  \label{eq:nkseq}
  \left( - \sum_{\alpha} \frac{\nabla_{\alpha}^2}{2 M_{\alpha}}
      + V_{S,n}(\nc) - \epsilon_n \right)
    \chi(\nc \ncs) &=& 0 \\
  \label{eq:ekseq}
  \left( -\frac{\nabla^2}{2} + v_{S,e}(\br) - \epsilon_{e,j}
  \right) \p_j(\br) &=& 0 .
\end{eqnarray}
By virtue of the MCHK theorem applied to the auxiliary system,
the effective potentials $V_{S,n}(\nc)$ and $v_{S,e}(\br)$
are uniquely determined by the ground-state
densities $\{\Gamma_0,\rho_0\}$, once their existence is assumed.
They are given by
\begin{eqnarray}
  \label{eq:vsn}
  V_{S,n}(\nc) &=& W_{nn}(\nc) + 
  \left.\frac{\delta E_{\uhxc}\den}{\delta \Gamma(\nc)} 
    \right|_{\Gamma_0,\rho_0} \\
  \label{eq:vse}
  v_{S,e}(\br) &=&   
   \left. \frac{\delta E_{\uhxc}\den}{\delta \rho(\br)}
  \right|_{\Gamma_0,\rho_0} .
\end{eqnarray}
In this procedure we require the nuclear wavefunction $\chi$ to have the same
symmetry-properties under the interchange of nuclei of the same type as the
exact wavefunction of the interacting system (this will also 
be required along the adiabatic connection to be discussed later in the paper). 
The last terms on the right-hand sides of Eqs.~(\ref{eq:vsn}) and
(\ref{eq:vse}) represent the potentials due to all non-trivial interactions of the system, 
i.e., they contain the Hartree-exchange-correlation (Hxc) effects of the electron-electron
and electron-nuclear interactions as well as mass-polarization and Coriolis
effects and the influence of the true external potentials $\hat{U}$. 
As seen in Eqs.~(\ref{eq:vsn}) and (\ref{eq:vse}), these potentials are given as functional
derivatives of the UHxc energy functional defined by
\begin{equation}
  \label{eq:euhxcdef}
  E_{\uhxc}\den := F\den - T_{S,n}[\Gamma] - T_{S,e}[\rho].
\end{equation}
This quantity represents the central quantity of the MCDFT and 
contains all many-body effects except the purely nuclear correlations.
We note that, in the case of vanishing external potentials $\hat{U} \equiv 0$, the
nuclear effective potentials $V_{S,n}(\nc)$ and the conjugated density, i.e., the
nuclear density matrix $\Gamma(\nc)$ are invariant under translations. 
Therefore, the nuclear center-of-mass can be separated off in Eq.~(\ref{eq:nkseq}), reducing the
number of degrees of freedom by three. We will illustrate this procedure in 
our applications later. 

In order to derive the above representations of the effective potentials,
we consider the energy functional of the auxiliary system introduced above:
\begin{eqnarray}
  \label{eq:estot}
  E_S\den &=& T_{S,n}[\Gamma] +  T_{S,e}[\rho]
  + \int d^{N_n}\brc \;  \Gamma(\nc)  V_{S,n}(\nc) \nonumber \\
  &+&  \int d\br \; \rho(\br) v_{S,e}(\br).
\end{eqnarray}
As noted before, the nuclear-nuclear interaction $\hat{W}_{nn}$
is included in the 'external' potential $V_{S,n}(\nc)$.
The noninteracting kinetic-energy functional $T_{S,e}[\rho]$ is the
one familiar from purely electronic DFT,
\begin{equation}
  \label{eq:ekine}
  T_{S,e}[\rho] = \min_{\Phi \to \rho} \langle \Phi| \hat{T}_e | \Phi 
\rangle ,
\end{equation}
where the minimization is over all electronic Slater determinants $\Phi$
yielding $\rho$.
Similarly, the nuclear kinetic-energy functional is given by
\begin{equation}
  \label{eq:ekinn}
  T_{S,n}[\Gamma] = \min_{\chi \to \Gamma}
  \langle \chi| \hat{T}_n | \chi \rangle .
\end{equation}
In contrast to the electronic wavefunction $\Phi$, the nuclear wavefunction
$\chi$ is not a Slater determinant, but a correlated many-body wavefunction,
since it minimizes $\hat{T}_n$ under the constraint of generating
the diagonal of the nuclear $N_n$-particle density matrix.
We note that, although $\chi$ is an interacting many-body wavefunction,
$T_{S,n}$ is not the interacting nuclear kinetic-energy functional
$T_n\den = \langle \Psi\den | \hat{T}_n | \Psi\den \rangle$, since
$\Psi\den$ minimizes $\langle \Psi | \hat{T} + \hat{W} + \hat{U}| \Psi \rangle$
(for given densities $\{\Gamma,\rho\}$), therefore including
all electron-nuclei interactions as well as mass-polarization
and Coriolis couplings.
Assuming the densities $\{\Gamma,\rho\}$ to be
{\it noninteracting $\{V_n,v_e\}$-representable},
the minimizing states of (\ref{eq:ekine}) and (\ref{eq:ekinn}), i.e.,
the states minimizing the kinetic energy for given $\{\Gamma,\rho\}$,
are obtained from Eqs.~(\ref{eq:nkseq}) and (\ref{eq:ekseq}) with
the potentials uniquely determined by the Euler equations following from
(\ref{eq:estot}):
\begin{eqnarray}
  \label{eq:eulers1}
  \left. \frac{\delta T_{S,n}[\tilde{\Gamma}]}{\delta \tilde{\Gamma}
      (\nc)} \right|_{\Gamma} + V_{S,n}(\nc)  &=& 0\\
  \label{eq:eulers2}
  \left. \frac{\delta T_{S,e}[\tilde{\rho}]}{\delta \tilde{\rho}(\br)}
  \right|_\rho  +  v_{S,e}(\br)  &=& 0\:.
\end{eqnarray}
Returning to the interacting problem, we decompose the functional
$F\den$ according to Eq.~(\ref{eq:euhxcdef}).
Employing this definition in the variational equations (\ref{eq:etot5})
and (\ref{eq:etot6}) of the interacting problem and comparing
them to the Euler equations (\ref{eq:eulers1})
and (\ref{eq:eulers2}), we find that the effective potentials which
reproduce the exact densities from the auxiliary system are indeed
given by Eqs.~(\ref{eq:vsn}) and (\ref{eq:vse}). 

Eqs.~(\ref{eq:ksnden})-(\ref{eq:ekseq}), (\ref{eq:vsn}) and (\ref{eq:vse})
constitute the MCKS system.
Since the effective potentials depend on both densities,
the MCKS equations (\ref{eq:nkseq}) and (\ref{eq:ekseq}) are coupled,
reflecting the mutual influence of electrons and nuclei on each other,
and have to be solved self-consistently.
We emphasize that Eq.~(\ref{eq:ekseq}), although
similar to the usual electronic KS equation, does not
parametrically depend on the nuclear configuration.
Instead, the information on the nuclear distribution is already
included through the functional dependence on $\Gamma$.
Considering the nuclear MCKS equation (\ref{eq:nkseq}),
we again realize its similarity with the conventional
nuclear BO equation. 
Yet, no BO approximation has been used to derive Eq.~(\ref{eq:nkseq}).
In contrast, since the MCKS scheme provides the exact ground
state, all non-BO effects are, in principle, included.
Whether or not the non-BO effects are reproduced in practical
applications depends, of course, on the quality of the approximations
employed for $E_{\uhxc}\den$. We also note that in the absence of
external potentials the potential $V_{S,n} (\nc)$ has the same 
symmetry properties as the BO-energy surface under rotations
and translations, i.e.
\begin{equation}
V_{S,n} (O \nc + \mathbf{a}) = V_{S,n} (\nc)
\end{equation}
The way it will transform under interchange of like nuclei will
depend on the way we choose the body-fixed frame. It is also important
to realize that when solving the
nuclear equation~(\ref{eq:nkseq}) we must look for the solution
$\chi (\nc \ncs)$ with the lowest energy under the constraint that
it has the correct symmetry under interchange of nuclear space-spin coordinates
, i.e. the symmetry that was imposed by the constrained-search.
Like the nuclear BO equation, the nuclear equation (\ref{eq:nkseq})
is still a many-body equation.
Therefore, its solution will, in general, be rather complicated
and further simplifications are highly desirable.
Typically, one first splits off the nuclear center-of-mass motion
and the global rotations of the molecule. Then the remaining
nuclear degrees of freedom are transformed to normal
coordinates, in terms of which the problem is treated in a
harmonic approximation, possibly including anharmonic effects in
a mean-field fashion~\cite{Gerber}. 
However, due to the generality of the method, different treatments
appropriate for different physical situations can be used.

\section{Analysis of the Functionals}
\label{sec:analysis}
\subsection{Decomposition of the Energy Functional}
\label{ssec:decompo}

In the last section, the foundations of the MCDFT were developed.
We derived a formally exact scheme, which provides a way
to calculate ground-state properties of MC systems.
For any practical application, the functional $E_{\uhxc}$ needs to be 
approximated.
In order to gain more insight in the construction of such an approximation, 
this section discusses a number of rigorous properties of this functional.

Following \cite{Gidopoulos:98}, we start out by decomposing the UHxc 
energy functional (\ref{eq:euhxcdef}) in parts associated with its various 
interactions.
To this end, we define the following quantities:
\begin{eqnarray}
  \label{eq:fedef}
  F^e[\rho] &:=& \min_{\psi \to \rho} \langle
  \psi | \hat{T}_e + \hat{W}_{ee} | \psi \rangle \\
  \label{eq:fendef}
  F^{en}[\Gamma, \rho] &:=& \min_{\psi \to \Gamma, \rho} \langle
  \psi | \hat{T}_n + \hat{T}_e  + \hat{W} | \psi \rangle \\
  \label{eq:tmpcdef}
  T_{\rm MPC}[\Gamma, \rho] &:=& \min_{\psi \to \Gamma, \rho} \langle
  \psi | \hat{T}_n  + \hat{T}_e  + \hat{T}_{\rm MPC}  +  \hat{W} | \psi 
  \rangle \nonumber \\
   && - F^{en} [ \Gamma, \rho ] \\
  \label{eq:eudef}
  U_{\hxc}[\Gamma, \rho] &:=& F[\Gamma,\rho] - \min_{\psi \to \Gamma, 
  \rho} \langle
  \psi | \hat{T} + \hat{W} | \psi \rangle .
\end{eqnarray}
The first term represents the electronic functional
which, by construction, is {\it identical} to the functional $F_{LL}[\rho]$
of standard electronic DFT, first introduced in \cite{Levy:79}.
Usually, this quantity is split according to
\begin{equation}
F^e[\rho] = T_{S,e}[\rho] + E^e_{\rm H}[\rho] + E^e_{\rm xc}[\rho]
\label{eq:elecF}
\end{equation}
where $E^e_{\rm H}$ is the electronic Hartree functional
\begin{equation}
E^e_{\rm H} [ \rho ] := \frac{1}{2} \int d\br d\br' 
\frac{\rho (\br) \rho (\br')}{|\br -\br'|}  
\end{equation}
and where the electronic exchange-correlation functional $E^e_{\rm xc}$ 
is defined by Eq.(\ref{eq:elecF}).
In contrast to $F^e$, the second functional $F^{en}$ also includes the 
nuclear kinetic energy as well as the electron-nuclear interaction, but 
still neglects mass-polarization and Coriolis effects and the influence of 
the external potential $\hat{U}$.
As discussed later on, the functional $F^{en}$ thus includes in particular the
effects arising from the electron-nuclear correlation.
The first term on the right-hand side of Eq.~(\ref{eq:tmpcdef}) additionally 
contains the mass-polarization and Coriolis terms of the kinetic-energy 
operator.
Therefore, the difference between $\min \langle \psi | \hat{T} + \hat{W} | 
\psi \rangle$ and $F^{en}$ is responsible for mass-polarization and 
Coriolis effects and thus denoted by $T_{\rm MPC}$.
Similarily, the last term denoted by $U_{\hxc}$ takes care of all effects 
introduced by the true external potentials $\hat{U}$.
Consequently, if no true external fields are applied to the system, 
$U_{\hxc}$ vanishes identically.

Inserting Eqs.~(\ref{eq:fedef})-(\ref{eq:eudef}) into Eq.~(\ref{eq:euhxcdef})
leads to
\begin{eqnarray}
  \label{eq:ehxcdecomp}
  E_{\uhxc} \den &=& E_{\rm H}^e [ \rho ]+ E_{\rm xc}^e[ \rho ] + \ehc \den \nonumber \\
  &+&
  T_{\rm MPC}\den + U_{\hxc}\den,
\end{eqnarray}
where
\begin{equation}
  \label{eq:ehcdef}
  \ehc\den := F^{en}\den - T_{S,n}[\Gamma] - F^e[\rho] .
\end{equation}
Eq.~(\ref{eq:ehxcdecomp}) provides a decomposition of the Hxc energy
functional in its natural contributions.
The first part, given by $E_{\rm H}^e$ and $E_{\rm xc}^e$, describes the
Coulomb interactions among the electrons.
It is important to note that these functionals are, by construction,
{\it identical} to the ones familiar from standard electronic DFT.
The electron-electron interaction can therefore be treated in the
familiar way, namely by using the widely investigated and
highly successful approximations for the electronic xc energy functional
$E_{\rm xc}^e[\rho]$.
The last term of Eq.~(\ref{eq:ehxcdecomp}) was constructed to
incorporate all effects arising from the presence of true external potentials
$\hat{U}$.
As already mentioned above, these terms are not of a
single particle form in the transformed coordinate system and have to be treated 
similarly to the interaction terms.
The functional $U_{\hxc}\den$ provides a means of dealing with these effects.
Similarly, the fourth term of Eq.~(\ref{eq:ehxcdecomp}) incorporates 
all effects due to the mass-polarization and Coriolis terms.
At least for ground-state properties, this term is expected to be
unimportant and can be neglected in most situations.
If such effects are, on the other hand, important in a given physical
situation, they can, in principle, be included in the calculation by
taking the functional $T_{\rm MPC}\den$ explicitly into account.
Finally, the term $\ehc\den$ contains all effects due to the electron-nuclear
interaction.
Its analysis will be continued in the next section.

The decomposition (\ref{eq:ehxcdecomp}) of the energy functional $E_{\uhxc}$
is obviously not unique.
However, the charme of the above prescription lies in the fact that, firstly, 
parts like the purely electronic functionals are already well known such 
that one can rely on existing approximations for these functional.
Secondly, the functionals (\ref{eq:fendef})-(\ref{eq:eudef}) contain, 
by their very construction, just the effect of one specifically chosen 
interaction.
This, in particular, guarantees that the functionals $E^e_{\hxc}[\rho]$, 
$E^{en}_{\hc}\den$ and 
$T_{\rm MPC}\den$ are {\em universal} in the sense that they do not depend on 
the external potentials and can therefore be employed for all systems independent 
of the applied external fields.
All effects arising from the external potentials are subsumed in 
$U_{\hxc}\den$.

\subsection{The electron-nuclear energy functional $E^{en}_{\hc}$}
\label{ssec:een}

Using the decomposition (\ref{eq:ehxcdecomp}), the well-studied
electronic Hxc energy functional as well as the -- 
at least for ground-state properties -- presumably negligible 
mass-polarization and Coriolis contribution were separated off in the
functional $E_{\uhxc}$.
In this section, we discuss the functional $E^{en}_{\hc}\den$ which
contains the many-body effects due to the electron-nuclear interaction.
We will derive an equation for this functional that is of a 
suitable form to be used in our approximations later.
To do this we will use the familiar coupling constant integration
technique of standard density-functional theory.
To begin with, we consider the Hamiltonian
\begin{equation}
  \label{eq:hlambda}
  \h_{\lambda} = \hat{T}_n + \hat{T}_e + \hat{W}_{ee}
  + \lambda \hat{W}_{en} + \hat{V}_{n,\lambda} + \hat{V}_{e,\lambda},
\end{equation}
where a non-negative coupling constant $\lambda$, scaling
the electron-nuclear interaction, has been introduced.
As usual, the potential $\hat{V}_{\lambda} = \hat{V}_{n,\lambda} +
\hat{V}_{e,\lambda}$ is chosen such that the densities remain fixed:
$\Gamma_{\lambda} = \Gamma$ and $\rho_{\lambda} = \rho$, independent of
the coupling constant $\lambda$.
Employing the coupling-constant integration
technique \cite{HarrisJones:74,LangrethPerdew:75,GunnarssonLundqvist:76}
adapted to the electron-nuclear interaction \cite{Gidopoulos:98},
the electron-nuclear Hc energy functional (\ref{eq:ehcdef})
is rewritten as
\begin{eqnarray}
 \lefteqn{ \ehc\den} \nonumber \\ 
&=& \min_{\Psi^{\lambda} \to \Gamma,\rho}  \left. \langle
    \Psi^{\lambda} | \hat{T}_n + \hat{T}_e + \hat{W}_{ee}
    + \lambda \hat{W}_{en}  | \Psi^{\lambda} \rangle \right|_{\lambda=1}
  \nonumber \\ && \hspace{-3.6mm}
  - \min_{\Psi^{\lambda} \to \Gamma,\rho}  \left. \langle
    \Psi^{\lambda} | \hat{T}_n + \hat{T}_e + \hat{W}_{ee}
    + \lambda \hat{W}_{en}  | \Psi^{\lambda} \rangle \right|_{\lambda=0}
  \nonumber \\
  &=& \int_0^1 d\lambda \; \frac{\partial}{\partial \lambda} \;
  \langle \Psi^{{\rm min},\lambda}_{\Gamma,\rho}
  | \hat{T}_n + \hat{T}_e
  + \hat{W}_{ee} + \lambda \hat{W}_{en} |
  \Psi^{{\rm min},\lambda}_{\Gamma,\rho}  \rangle
  \nonumber \\
  &=& \int_0^1 d\lambda \; \langle \Psi^{{\rm min},\lambda}_{\Gamma,\rho}
  | \hat{W}_{en} | \Psi^{{\rm min},\lambda}_{\Gamma,\rho}  \rangle,
\label{eq:ehc}
\end{eqnarray}
where $\Psi^{{\rm min},\lambda}_{\Gamma,\rho}$ denotes the minimizing 
state of
$\langle \hat{T}_n + \hat{T}_e + \hat{W}_{ee}
+ \lambda \hat{W}_{en} \rangle$ generating the given densities
$\{\Gamma,\rho\}$, and a Hellmann-Feynman-type theorem was used in the last 
step.
Therefore, the electron-nuclear energy functional $\ehc$ is given by
\begin{eqnarray}
  \lefteqn{\ehc\den} \label{eq:ehc2} \\
 &=& \int d^{N_n}\brc \; \Gamma(\nc)
  \int d\br \; W_{en}(\nc,\br) \;
  \bar{\gamma}^{\rm min}[\Gamma,\rho](\br|\nc) \nonumber,
\end{eqnarray}
where
\begin{eqnarray}
  \label{eq:wendef}
  W_{en}(\nc,\br) &=& - \sum_{\alpha} \frac{Z_{\alpha}}
  {|\mathcal{R}^{-1}\br-\brc_{\alpha} + \brc_{CMN}|} \nonumber \\
 &=& 
   - \sum_{\alpha} \frac{Z_{\alpha}}
  {| \br- \mathcal{R} (\brc_{\alpha} - \brc_{CMN}) |} .
\end{eqnarray}
The electronic conditional density $\gamma$ is defined by
\begin{equation}
  \label{eq:econddendef}
  \gamma(\br|\nc) := N_e \sum_{\sigma,s} \int d^{N_e-1}\br \; 
	| \Psi(\ec \ecs,\nc \ncs) |^2 / \Gamma(\nc) ,
\end{equation}
and $\bar{\gamma}$ represents the coupling-constant average of 
$\gamma$. The conditional density satisfies important sumrules that we will
use later for the construction of approximate functionals:
\begin{eqnarray}
N_e &=& \int d\br \; \gamma (\br | \nc) \qquad \forall \nc 
\label{eq:sumrule1} \\
\rho (\br) &=& \int d\nc \; \Gamma (\nc) \gamma (\br | \nc)
\label{eq:sumrule2}
\end{eqnarray}
By virtue of Eq.~(\ref{eq:ehc2}), the Hc energy can be interpreted
as the electrostatic interaction energy of the (coupling-constant
averaged) electronic density for a fixed nuclear configuration
with the point charges of the corresponding nuclei, averaged over
the nuclear distribution. This interpretation will play an important role
in our later development of approximate functionals.

In order to gain further insight into the electron-nuclear Hc 
energy functional, we now establish a connection
between the MCDFT scheme and the conventional BO method which provides a highly
successful treatment of electron-nuclear correlation.
To that end, we decompose the total wavefunction into an adiabatic product
according to
\begin{equation}
  \label{eq:adiabatic}
  \Psi^{\rm min}_{\Gamma,\rho}(\ec \ecs,\nc \ncs) = 
 \chi(\nc \ncs) \; \Xi\den(\ec \ecs|\nc \ncs),
\end{equation}
where $\chi$ is the nuclear wavefunction generating the nuclear
density matrix $\Gamma$ and
$\Xi\den$ is an electronic state normalized to one for every 
nuclear configuration $\nc \ncs$ :
\begin{equation}
\sum_{\sigma} \int d^{N_e} \br | \Xi\den(\ec \ecs|\nc \ncs )|^2 =1 .
\label{eq:xinorm}
\end{equation}
We note that the decomposition~(\ref{eq:adiabatic})
is actually an exact representation of the correlated
electron-nuclear wave function~\cite{Hunter0,Hunter1,Hunter2,Hunter3,CzubWolniewicz,CassamChenai} and that the factors
$\chi$ and $\Xi$ are unique~\cite{GidopoulosGross:condmat}
up to within an $\nc \ncs$-dependent phasefactor.
However, it is important to note that the electronic state
$\Xi$ is {\it not} identical to the usual electronic BO state.
Even if non-BO effects were neglected, $\Xi$ would not be identical to
the electronic BO state since $\chi$ and $\Xi$ are required to reproduce
a given set of densities ($\Gamma,\rho$)
(the two electronic wavefunctions only become equivalent, 
if $\Xi[\Gamma,\rho]$ is evaluated
at the BO densities ($\Gamma,\rho$)=($\Gamma^{\rm BO},\rho^{\rm BO}$)).
Instead, $\Xi\den$ is expanded according to
\begin{equation}
  \label{eq:boexpand}
  \Xi\den(\ec \ecs|\nc \ncs) = \sum_k a_k\den(\nc \ncs) \;
  \Xi^{\rm BO}_{\nc,k}(\ec \ecs),
\end{equation}
where $\{\Xi^{\rm BO}_{\nc,k}\}$ denotes a complete set of (BO)
eigenfunctions corresponding to the electronic (clamped-nuclei) Hamiltonian
$\hat{H}_e := \hat{T}_e + \hat{W}_{ee} +\hat{W}_{en}$.
Employing Eq.~(\ref{eq:adiabatic}) together with (\ref{eq:xinorm})
and assuming that $\Xi [\Gamma,\rho]$ is real,
the electron-nuclear Hc energy functional is given from Eq.(\ref{eq:ehcdef}) 
by
\begin{eqnarray}
\lefteqn{\ehc\den} \\    &=&\langle \Psi^{\rm min}_{\Gamma,\rho} |
	\hat{T}_n +  \hat{H}_e | \Psi^{\rm min}_{\Gamma,\rho} \rangle
 	- T_{S,n}[\Gamma]  - F^e[\rho] \nonumber \\ 
&=& \sum_{s} \int d^{N_n} \brc \; | \chi (\nc \ncs)|^2  \langle \Xi \den | \hat{T}_n + \hat{H}_e | \Xi \den \rangle_e
- F^e[ \rho] \nonumber
\end{eqnarray}
where the index ``e" at the bracket indicates that the integration is over electronic
coordinates only.
Using Eq.(\ref{eq:boexpand}) we then obtain
\begin{eqnarray}
 \lefteqn{ \ehc\den} \label{eq:ebo3} \\ 
  &=&	\sum_s \int d^{N_n}\brc \; |\chi (\nc \ncs)|^2
  \sum_{k,l}  \Big\{
  \left| a_k\den(\nc \ncs) \right|^2 \cdot
  \epsilon^{\rm BO}_k(\nc) \delta_{k,l} \nonumber \\
  &+& a_k^{\star}\den(\nc \ncs) \,
  \langle  \Xi^{\rm BO}_k| \hat{T}_n | \Xi^{\rm BO}_l  \rangle_e\,
  a_l\den(\nc \ncs ) \Big\}   - F^e[\rho] , \nonumber 
\end{eqnarray}
where
\begin{equation}
  \label{eq:bopes}
  \epsilon^{\rm BO}_k(\nc) = \langle
  \Xi^{\rm BO}_k | \hat{H}_e | \Xi^{\rm BO}_k  \rangle_e
\end{equation}
represents the $k$th BO potential-energy surface (PES)
and the index ``e'' at the bracket indicates that the integration is
over electronic coordinates only.
On the basis of Eq.~(\ref{eq:ebo3}), one can interpret $\ehc$ as
the potential energy of the nuclei,
where the nuclear distribution lies in a potential
hypersurface, which is composed of adiabatic BO-PES weighted
with the coefficients $a_k\den$ as well as nonadiabatic corrections to it.
Of course, the coefficients $a_k$ and their functional dependence on the
set of densities $\den$ is unknown at this point.
However, Eq.~(\ref{eq:ebo3}) helps us in gaining a better understanding
of the electron-nuclear Hc energy functional and establishes an -- at
least -- formal link to the BO scheme which is further exploited when
the effective potentials are discussed later on.

\subsection{Concerning the true external potentials}
\label{ssec:truext}

Similar to the techniques employed in the last section, the 
coupling-constant integration can be employed to derive an expression
for the functional $U_{\hxc}\den$ which subsumes the many-body
effects arising from the true external field.
In analogy to Eq.~(\ref{eq:ehc2}), one obtains
\begin{eqnarray}
  \label{eq:euhxc}
\lefteqn{ U_{\hxc} \den } \\ 
&=& \int d^{N_n}\brc \; \Gamma(\nc)
  \int d\br \; U_{\rm ext}(\nc,\br) \;
  \bar{\gamma}^{\rm min'}[\Gamma,\rho](\br|\nc) \
\nonumber,
\end{eqnarray}
where $\bar{\gamma}^{\rm min'}$ again denotes the coupling-constant
average with respect to coupling constant $\mu$ of the conditional density $\gamma^{{\rm min'},\mu}\den$
corresponding to the states $\Psi^{{\rm min'},\mu}_{\Gamma,\rho}$ 
minimizing $\langle \hat{T} + \hat{W} + \mu \hat{U} \rangle$.
We further defined $U_{\rm ext} (\nc,\br)$ to be 
\begin{equation}
U_{\rm ext} (\nc, \br) :=  \frac{1}{N_e}  U_{{\rm ext},n} (\nc)
 + u_{{\rm ext},e} (\mathcal{R}^{-1} \br + \brc_{CMN})
\end{equation}
It has to be noted that the conditional densities appearing in 
Eqs.~(\ref{eq:ehc2}) and (\ref{eq:euhxc}) are {\it not} identical since the
corresponding states $\Psi^{{\rm min},\lambda}_{\Gamma,\rho}$ and
$\Psi^{{\rm min'},\mu}_{\Gamma,\rho}$ minimize different expressions.
This is a direct consequence of the definitions chosen in 
Eqs.~(\ref{eq:fedef})-(\ref{eq:eudef}).
In particular, this choice guarantees that $E^e_{\hxc}$, $T_{\rm MPC}$, 
and $\ehc$ are {\em independent} of the true external potential $U_{\rm ext}$,
i.e., these functionals are universal;
all effects stemming from $U_{\rm ext}$ are contained exclusively
in the functional $U_{\hxc}$.

By virtue of the above discussion, the influence of the true external potential
has to be treated similar to an interaction.
As already mentioned above, this complication is an immediate consequence of
the necessity to transform to an internal reference system for the 
formulation of the MCDFT scheme.
Of course, in the numerous cases discussing the properties of {\em isolated}
systems, $\hat{U}$ vanishes and the MCDFT formalism reduces
to the one given in \cite{KreibichGross:2001}.
If, on the other hand, a true external potential is applied to the system,
approximations for $U_{\hxc}$ are needed.
In the simplest case, the electronic conditional density $\gamma$ 
is replaced by the electronic density $\rho$, leading to a Hartree-type
approximation for $U_{\hxc}$.
Such an approximation will be especially valid in the case of well
localized nuclei, as discussed later on.

\subsection{Analysis of the Effective Potentials}
\label{ssec:anapot}

In the last sections, the Hxc energy functional of Eq.(\ref{eq:euhxcdef}) was discussed.
According to Eqs.~(\ref{eq:vsn}) and (\ref{eq:vse}), this quantity gives
rise to the many-body contributions of the effective MCKS potentials.
Explicitly, the UHxc potentials are given by
\begin{eqnarray}
  \label{eq:vnhxc}
  V_{\uhxc}\den(\nc) &=&
  \frac{\delta E_{\uhxc}\den}{\delta \Gamma(\nc)} \\
  \label{eq:vehxc}
  v_{\uhxc}\den(\br) &=& \frac{\delta E_{\uhxc}\den}{\delta \rho(\br)} .
\end{eqnarray}
Employing Eq.~(\ref{eq:ehxcdecomp}), the potentials can also be decomposed 
into the parts associated with the different interactions, yielding
\begin{eqnarray}
  \label{eq:vnhxc2}
  V_{\uhxc}\den(\nc) &=& V^{U}_{\hxc}\den(\nc) +
  V^{en}_{\hxc}\den(\nc) \nonumber \\
    &+& V_{\rm MPC}\den(\nc)\\
  \label{eq:vehxc2}
  v_{\uhxc}\den(\br) &=& v^{U}_{\hxc}\den(\br) +
  v^e_{\rm H}[\rho](\br) + v^e_{\rm xc}[\rho](\br) \nonumber \\
&+& v^{en}_{\hc}\den(\br) + v_{\rm MPC}\den(\br),
\end{eqnarray}
where the various potential terms on the right-hand sides of the 
above equations are defined in analogy to (\ref{eq:vnhxc})
and (\ref{eq:vehxc}):
The first terms on the right-hand side of Eqs.~(\ref{eq:vnhxc}) and
(\ref{eq:vehxc}) represent the influence of the true external potential
$\hat{U}$ and correspond to the derivatives of $U_{\hxc}$ in Eq.(\ref{eq:ehxcdecomp}).
Since the electron-electron interaction is treated employing the well-known
Hxc energy functional $E^e_{\hxc}[\rho]$ from standard electronic DFT,
the corresponding potentials $v^e_{\rm H}$ and $v^e_{\rm xc}$ are also
identical to the familiar electronic Hartree and xc potentials.
Furthermore, as for the energy functional, the potentials arising from the
mass-polarization and Coriolis effects are not expected to contribute
significantly -- at least for ground-state properties.
In the following we will concentrate on the Hxc potentials arising from the
electron-nuclear energy functional $\ehc$.

To start with, we consider the nuclear Hc potential,
defined by
\begin{equation}
  \label{eq:vnhcen}
  V^{en}_{\hc}\den(\nc) =
  \frac{\delta E^{en}_{\hc}\den}{\delta \Gamma(\nc)} .
\end{equation}
Employing the representation  of $\ehc$ in terms
of the coupling-constant averaged conditional density, Eq.~(\ref{eq:ehc2}), 
the nuclear potential can be split in two parts,
\begin{equation}
  \label{eq:vnsplit}
  V^{en}_{\hc}\den(\nc) = V^{en}_{\rm cond}\den(\nc)+
  V^{en}_{\rm c,rsp}\den(\nc)
\end{equation}
where
\begin{equation}
  \label{eq:vncond}
  V^{en}_{\rm cond}\den(\nc) :=
  \int d\br \; W_{en}(\nc,\br) \;
  \bar{\gamma}^{\rm min}[\Gamma,\rho](\br|\nc)
\end{equation}
is the electrostatic potential due to the electronic conditional
density and
\begin{eqnarray}
  \label{eq:vnrsp}
   V^{en}_{\rm c,rsp}\den(\nc) := \int d^{N_n}\brc' \; \Gamma(\nc') \nonumber \\
    \times \int d\br \; W_{en}(\nc',\br)
  \frac{\delta \bar{\gamma}^{\rm min}[\Gamma,\rho](\br|\nc')}
  {\delta\Gamma(\nc)}
\end{eqnarray}
defines a response-type contribution to the electron-nuclear correlation
potential.
We note that the conditional potential $V^{en}_{\rm cond}$
completely determines the electron-nuclear Hc energy:
\begin{equation}
  \label{eq:hxcfromvcond}
  \ehc\den = \int d^{N_n}\brc \; \Gamma(\nc)
  V^{en}_{\rm cond}\den(\nc) .
\end{equation}
In the following analysis we restrict ourselves to situations where
the full electron-nuclear wave function  $\Psi^{\rm min}_{\Gamma,\rho}$
can be factorized into a nuclear spin function times a remainder
not depending on $\ncs$. This is exactly true, e.g., for diatomic 
molecules or when the nuclei are spin-zero bosons. In many
other cases, this factorization represents a good approximation.
Under these circumstances, the wave function $\Xi$ in
Eq.(\ref{eq:adiabatic}) can be chosen to be independent 
of $\ncs$, and likewise the expansion coefficients  $a_k$
in Eq.(\ref{eq:boexpand}), so that Eq.(\ref{eq:ebo3})
reduces to
\begin{eqnarray}
 \lefteqn{ \ehc\den} \label{eq:ebo3old} \\ 
  &=&	\int d^{N_n}\brc \; \Gamma (\nc)
  \sum_{k,l}  \Big\{
  \left| a_k\den(\nc) \right|^2 \cdot
  \epsilon^{\rm BO}_k(\nc) \delta_{k,l} \nonumber \\
  &+& a_k^{\star}\den(\nc) \,
  \langle  \Xi^{\rm BO}_k| \hat{T}_n | \Xi^{\rm BO}_l  \rangle_e\,
  a_l\den(\nc ) \Big\}   - F^e[\rho] , \nonumber 
\end{eqnarray}
Comparing Eq.(\ref{eq:hxcfromvcond}) and (\ref{eq:ebo3old}), the
conditional potential
(\ref{eq:vncond}) can be expressed in terms of the BO-PES:
\begin{eqnarray}
  \label{eq:vncondbopes}
 \lefteqn{ V^{en}_{\rm cond}\den(\nc)
  =  \sum_k \left| a_k\den(\nc) \right|^2  \epsilon^{\rm BO}_k(\nc)}
  \nonumber \\&+& \sum_{k,l} a_k^{\star}\den(\nc) \,
  \langle  \Xi^{\rm BO}_k| \hat{T}_n | \Xi^{\rm BO}_l  \rangle_e\,
  a_l\den(\nc)  \nonumber \\  
&-& F^e[\rho] .
\end{eqnarray}
This equation provides a useful tool to interpret
the effective nuclear MCKS potential:
The first term in (\ref{eq:vncondbopes}) is a weighted sum over different
adiabatic BO-PES, whereas the
second one describes adiabatic and nonadiabatic corrections to it.
The last term in equation (\ref{eq:vncondbopes}), $F^e[\rho]$, just
yields a constant shift and is included in the potential to maintain
the same zero-energy level within the BO and MCKS schemes.
Considering the case that the BO approximation accurately describes
a specific system, we realize that, in the first sum, 
only the lowest coefficient
$a_0$ survives and the second sum is negligible, provided the potential is
evaluated at the ground-state densities.
Therefore, $V^{en}_{\rm Hc}(\nc) \approx \epsilon^{\rm BO}_0(\nc)$,
and the nuclear MCKS equation reduces to the nuclear
BO equation in the limit considered here.
We emphasize, however, that the way to evaluate this potential differs
in the MCKS and BO methods.
Whereas, in the latter, an electronic equation has to be solved
for each nuclear configuration,
the MCKS potential is determined by the {\it functional}
derivative $\delta E^{en}_{\rm Hc}\den / \delta \Gamma$.
Inserting the ground-state densities then yields a potential
which, as a {\it function} of $\nc$, is very close to the BO potential
(in the case discussed here).
Furthermore, we can conclude that the response part of the nuclear
potential, Eq.~(\ref{eq:vnrsp}), has negligible influence for such systems.
If, on the other hand, nonadiabatic effects -- e.g. close to level
crossings -- are encountered, the coefficients $a_k$ in
Eq.~(\ref{eq:vncondbopes}) will, as a function of the nuclear
configuration, achieve a natural diabatization.
One should also note that the electronic wavefunction
$\Xi$ is, in general, complex at points of degeneracy.
Therefore, one obtains another contribution to the nuclear
potential, which is responsible for Berry-phase effects
\cite{Berry:84} \footnote{
In the formalism presented here, the Berry-phase effects are assumed
to be representable by a scalar nuclear potential.
In view of the connection of the Berry phase to a vector potential
in the nuclear Schr\"odinger equation \cite{ShapereWilczek:89},
a multicomponent current-density
description appears more appropriate to treat these effects.}.
In addition, the response part of the nuclear potential might
contribute appreciably.
In summary, Eq.~(\ref{eq:vncondbopes}) shows that the (exact)
nuclear effective potential reduces to the lowest-energy BO-PES,
if nonadiabatic contributions can be neglected, but also contains
in principle all non-BO effects.
Whether or not they can be recovered in an actual application crucially
depends, of course,
on the level of sophistication of the approximation used for $\ehc$.

Employing again Eq.~(\ref{eq:ehc2}), the electronic potential 
due to the electron-nuclear interaction,  defined by
\begin{equation}
  \label{eq:vehcdef}
  v^{en}_{\hc}\den(\br) := \frac{\delta \ehc\den}{\delta \rho(\br)},
\end{equation}
is given by:
\begin{eqnarray}
 \lefteqn{ v^{en}_{\hc}\den(\br) = \int d^{N_n}\brc \; \Gamma(\nc) } \nonumber \\
  &\times& \int d\br' W_{en}(\nc,\br')
  \frac{\delta \bar{\gamma}^{\rm min}[\Gamma,\rho](\br'|\nc)}
  {\delta \rho(\br)} .
\label{eq:veen}
\end{eqnarray}
This expression appears rather complicated to evaluate.
If, however, the nuclear probability distribution is sharply peaked
around an equilibrium geometry $\nceq$,
only configurations around $\nceq$ will substantially
contribute to the above integral.
Then,  the calculation of the
electronic Hxc potential simplifies to
\begin{equation}
  \label{eq:venapprox}
  v^{en}_{\hc}\den(\br) \approx \int d^{N_n}\brc  \;
  \Gamma(\nc) W_{en}(\nc, \br) .
\end{equation}
This potential represents the electrostatic (Hartree) potential due
to the nuclear charge distribution acting on the electrons.
Since the nuclear ground-state densities of many molecules are indeed
strongly localized functions -- in other words: the nuclei behave almost
classically -- we expect the Hartree approximation for the
electronic potential to be sufficiently accurate for such systems.
If, on the other hand,
the assumption of nicely localized nuclear densities breaks down,
one needs to incorporate correlation contribution to $v^{en}_{\hc}$ 
arising from the electron-nuclear interaction.

\subsection{The Limit of Classical (Point-Like) Nuclei}
\label{ssec:classlimit}

In this section, we investigate the limit of classical, i.e.,
perfectly localized nuclei.
Assuming identical zero spin nuclei for the ease of notation, the nuclear
density matrix reads
\begin{equation}
  \label{eq:gammaclass}
  \Gamma^{\rm class}(\brc_1...\brc_{N_n}) = \frac{1}{N_n !}
  \sum_{\mathcal{P}} \prod_{\alpha} \delta \left(
    \brc_{\mathcal{P}(\alpha)} - \brc_{\alpha,0} \right),
\end{equation}
where the sum is over all $N_n!$ permutations of the
nuclear coordinates and $\nc_0$ denotes the
positions where the nuclei are located. Note that by
this classical form of the density matrix we have broken
the translational and rotational symmetry of the
density matrix as presented in Eq.(\ref{eq:gam-trafo}).
In the following we investigate the consequences are of this form
for the diagonal of the nuclear density matrix.
First, we consider the electronic density.
In terms of the coupling-constant dependent conditional density,
it is given by
\begin{equation}
  \label{eq:nfromrho2}
  \rho(\br) = \int d^{N_n}\brc \; \Gamma(\nc)  \gamma^{\lambda}(\br|\nc).
\end{equation}
We recall that $\rho(\br)$ does not depend on the coupling constant
$\lambda$, since the external potentials are chosen such that
the densities remain unchanged.
Inserting Eq.~(\ref{eq:gammaclass}) into (\ref{eq:nfromrho2})
yields
\begin{equation}
  \label{eq:nclass1}
  \rho(\br) = \gamma^{\lambda}(\br|\nc_0) ,
\end{equation}
i.e., in the limit of classical nuclei, the electronic density is
identical to the conditional density evaluated at the positions
of the classical nuclei.
This quantity, in fact, serves as the basic variable
of standard electronic DFT employing the BO approximation:
$\rho^{\rm DFT,BO}(\br) = \gamma(\br|\nc_0)$.
We therefore conclude that the MCDFT presented here reduces to
the standard formulation of DFT in the limit of classical nuclei.

Inserting Eq.~(\ref{eq:nfromrho2}) into Eqs.~(\ref{eq:ehc2}) and
(\ref{eq:euhxc}), we readily obtain the expressions
for $\ehc$ and $U_{\hxc}$ in the classical limit:
\begin{eqnarray}
  \label{eq:ehcclass}
  \ehc [\Gamma^{\rm class},\rho] &=&
  \int d\br \;  \rho(\br)  W_{en}(\nc_0, \br) \\
  \label{eq:euhcclass}
  U_{\hxc} [\Gamma^{\rm class},\rho] &=&
  \int d\br \;  \rho(\br) U_{\rm ext}(\nc_0,\br) .
\end{eqnarray}
Thus, in the limit of classical nuclei, the Hxc energy functionals
reduce to the classical electrostatic (Hartree) interactions
and correlation contributions vanish \cite{Gidopoulos:98}.

The corresponding electronic potentials, 
following from Eqs.~(\ref{eq:ehcclass}) and (\ref{eq:euhcclass}) then read
\begin{eqnarray}
  \label{eq:vhceclass}
  v^{en}_{\hc}[\Gamma^{\rm class},\rho] = W_{en}(\nc_0,\br) \\
  \label{eq:vuhceclass}
  v^{U}_{\hxc}[\Gamma^{\rm class},\rho] = U_{\rm ext}(\nc_0,\br).
\end{eqnarray}
The first quantity is identical to the classical Coulomb field of the nuclei, 
whereas the second one describes the influence
of potentials applied externally to the system.
Both quantites together represent the ``external potential'' in
BO-based DFT, reflecting again its coincidence with the MCDFT in the limit 
of classical nuclei.
From this perspective, one also might consider Eq.~(\ref{eq:venapprox})
as the natural extension of Eq.~(\ref{eq:vhceclass}) to nuclear 
distributions which are localized but still exhibit a finite width.


\section{Applications}
\subsection{Diatomic molecules}
\label{sec:results}
Having discussed the foundation and some formal properties of the 
MCDFT, we proceed with the application of the theory to the case of 
isolated diatomic molecules.

However, as mentioned above, we treatment of the $N_n$-body nuclear
MCKS equation has to be discussed prior to actual applications:
Since ``true'' external potentials are absent, i.e., $\hat{U}
\equiv 0$ the system is translationally invariant.
Accordingly, the nuclear MCKS potential is required to behave as
$V_{S,n}(\brc_1,\brc_2) = V_{S,n}(\brc_1-\brc_2)$.
The nuclear equation has then the form
\begin{eqnarray}
\lefteqn{ \Big( -\frac{1}{2M_1} \nabla_{\brc_1}^2 -\frac{1}{2M_2} \nabla_{\brc_2}^2 } \nonumber \\
&&  +
V_{S,n} (\brc_1 -\brc_2) - \epsilon_n \Big) \chi (\brc_1 s_1, \brc_2 s_2) =0 
\end{eqnarray}
Then the nuclear CM motion can be separated off 
and the problem can be reformulated in terms of the internuclear 
separation $\brc := \brc_2 - \brc_1$.
The nuclear function $\chi$ has the general form
\begin{equation}
\chi (\brc_1 s_1, \brc_2 s_2) = \eta (\brc_{CM}) \xi (\brc_1 -\brc_2) 
\theta (s_1, s_2)
\end{equation}
where $\eta (\brc_{CM})$ is a plane wave state depending on the center-of-mass coordinate
$\brc_{CM} =(M_1 \brc_1 + M_2 \brc_2) /(M_1+M_2)$ and is explicitly given by
\begin{equation}
\eta (\brc_{CM}) = \frac{1}{\sqrt{V}} e^{i \mathbf{k} \cdot \brc_{CM}}
\end{equation}
where $V$ is the total volume of the system. The relative wavefunction $\xi (\brc_1-\brc_2)$
satisfies the nuclear MCKS equation
\begin{equation}
  \label{eq:ndikseq}
  \left( - \frac{\nabla_{\brc}^2}{2 \mu_n}
      + V_{S,n}(\brc) - \epsilon_n \right)
    \xi(\brc) = 0 ,
\end{equation}
where $\mu_n = M_1 M_2 / (M_1+M_2)$ denotes the reduced nuclear mass.
The function $\theta$ is a nuclear spin function depending on the type of
nuclear species. For instance, for the $H_2$ with two protons as nuclei the
nuclear ground state is that of para-hydrogen where the function $\theta$ is an antisymmetric spin
function and consequently the function $\xi$ must be even, i.e. $\xi (\brc) = \xi (-\brc)$, to preserve overall
anti-symmetry of the wavefunction under the interchange of the two protons.
For the density matrix $\Gamma$, one obtains after integrating out the spin function
\begin{equation}
  \label{eq:digamma}
  \Gamma(\brc_1,\brc_2) = \frac{1}{V}| \xi(\brc_1 - \brc_2)|^2 
  \equiv \frac{1}{V} \Gamma(\brc).
\end{equation}
Therefore, the diagonal of the nuclear density matrix, which we often - and somewhat
unprecisely - refer to as ``nuclear density'' is indeed a single-particle
quantity describing the probability of finding the two nuclei separated 
by $\brc$.\\
It remains to discuss the electronic coordinates.
For our diatomic molecule we determine the Euler angles by the requirement that 
the internuclear axis be parallel to the $z$-axis in the
body-fixed frame, i.e. $\mathcal{R} (\brc_1 - \brc_2)= R \mathbf{e}_z$
, where $R= | \brc_1 -\brc_2|$. For the special case of the diatomic
molecule only two Euler angles are needed to specify the rotation matrix
$\mathcal{R}$. The electronic coordinates in the body-fixed frame are then
obtained using Eq.(\ref{eq:rep}). With this transformation the electron-nuclear
attraction attains the form
\begin{equation}
  \label{eq:wendia}
  W_{en}(\brc,\br) = -\frac{Z_1}{|\br - \frac{M_2}{M_{\rm nuc}} R {\bf e}_z| }
    - \frac{Z_2}{|\br + \frac{M_1}{M_{\rm nuc}} R {\bf e}_z| }.
\end{equation}
As for any other DFT, explicit approximations have to be employed for the
energy functional $E_{\uhxc} \den$ of Eq.(\ref{eq:ehxcdecomp}).
Since no ``true'' external potentials are present in the case discussed
here, $U_{\hxc}\den$ vanishes identically.
Furthermore, following Sec.~\ref{ssec:decompo}, the purely
electronic part of the energy functional can be treated 
by using the familiar approximations for the electronic xc
energy functional $E_{\rm xc}^e[\rho]$.
For all systems containing more than one electron, we will employ the
well-known LDA approximation.
We emphasize that it is not the purpose of this work
to investigate new approximations for the {\it electronic}
xc energy functional.
Instead, we aim at an analysis of the previously not much studied
Hc energy functional arising from the electron-nuclear interaction.
To that end, we restrict ourselves to work within the LDA
approximation for $E_{\rm xc}^e[\rho]$; different -- and
more sophisticated -- approximations for the electronic xc energy 
functional would only result in some minor quantitative changes
of the analysis presented below.
Furthermore, we note that the mass-polarization and Coriolis effects
are not expected to contribute substantially to ground-state properties.
Therefore, only the diagonal part of the mass-polarization term which leads
to a reduced mass in the electronic MCKS equation is accounted for and the
remaining parts are neglected in all practical calculations.
With these assumptions the electronic Kohn-Sham equations for our problem
then attain the form
\begin{equation}
\left( -\frac{\nabla^2}{2 \mu_e} + v_{S,e} [ \Gamma, \rho ] (\br)  
 - \epsilon_{e,j} \right)
\varphi_{j} (\br) = 0
\label{eq:KSelec-dia}
\end{equation}
where
\begin{equation}
v_{S,e} [ \Gamma, \rho ] (\br)  = v_{\hc}^{en} [ \Gamma, \rho] (\br)  + 
v_{\rm H}^{e} [\rho] (\br) + v_{\rm xc}^e [\rho] (\br )
\label{eq:elec-dia}
\end{equation}
and $\mu_e = (M_1+M_2)/(M_1+M_2+1)$.
The approximations used for $v_{\hc}^{en}$ will be discussed
in the subsequent sections.
Within the same assumptions the effective potential in the
nuclear equations will be of the form
\begin{eqnarray}
V_{S,n} [\Gamma, \rho ] (\brc) = W_{ee} (\brc) + V_{\hc}^{en} [\Gamma, \rho](\brc)
\label{eq:nuc-dia}
\end{eqnarray}
where $W_{ee} (\brc)= Z_1 Z_2 / R$.\\
We finally note that the diatomic molecule is a particularly convenient case
for studying the nuclear effective potential $V_{S,n}$ since for this case 
for a given density $\Gamma (\brc)$ the potential $V_{S,n}$ is easily constructed
from inversion of Eq.(\ref{eq:ndikseq}):
\begin{equation}
V_{S,n}(\brc) = \epsilon_n + \frac{1}{2\mu_n} \frac{\nabla_{\brc}^2 \sqrt{\Gamma (\brc)}}
{\sqrt{ \Gamma (\brc)}}
\label{eq:KSinversion}
\end{equation}
We have done this for a one-dimensional model of the $H_2^+$
for which the exact nuclear density $\Gamma$ can be obtained by
numerical integration of the full many-body Schr\"{o}dinger equation.
In this way the exact nuclear potential $V_{S,n}$ for this
problem was obtained~\cite{Kreibich:thesis}. For this case we found 
the exact $V_{S,n}$ to be almost identical to the BO potential except for the case
when the nuclear masses were taken to be artificially small.
This illustrates the point discussed before: in situations where the BO approximation works well, 
the nuclear potential $V_{S,n}$ will be close to the BO potential.
For this reason the BO potentials of the $H_2$ and $H_2^+$ molecules
that we will study below will be a good reference to test our approximations for the
electron-nuclear correlation functional. Of course, having obtained
a good approximate functional its main field of applicability will
be cases where the BO approximation does not work well.\\ 
To conclude these introductory remarks, the
numerical implementation of the MCKS equations
(\ref{eq:ndikseq}) and (\ref{eq:KSelec-dia}) is briefly described.
Since Coriolis effects are neglected,
the effective nuclear potential is spherically
symmetric and the angular part can be treated analytically.
The remaining radial nuclear MCKS equation is numerically
solved on a one-dimensional grid.
Furthermore, we observe that the $z$-component of the electronic
angular momentum is a conserved quantity.
Hence, the electronic MCKS equation is rewritten in terms of cylindrical
coordinates.
For axial-symmetric electronic potentials, the angular part can
be integrated out and we are left with a two-dimensional problem.
The resulting Hamiltonian is then discretized on a
uniform rectangular grid and numerically diagonalized by employing 
the Lanczos algorithm \cite{Lanczos:50}.
Due to the use of finite uniform grids, the regions around the
nuclei are not sampled with high accuracy, leading to typical
discretization errors of about $0.1\%$ for the systems
discussed later in this section.
Both the nuclear and the electronic equation are solved
simultaneously until self-consistency is achieved.

\subsection{The Hartree Approximation for the electron-nuclear 
  energy functional}
\label{ssec:hartree}

It remains to find explicit approximations for the electron-nuclear
Hc energy functional $\ehc \den$.
In the simplest case, the electron-nuclear interaction
is approximated by the Hartree energy functional, defined by
\begin{eqnarray}
  \label{eq:enhartree}
  E_{\rm H}^{en}[\Gamma,\rho] &:=& \int d\brc_1 d\brc_2 d\br \; 
	\Gamma(\nc)W_{en}(\nc,\br) \rho(\br) \nonumber \\
	&=& \int d\brc d\br \; \Gamma(\brc) W_{en}(\brc,\br) \rho(\br) .
\end{eqnarray}
where in the second step we changed to relative $\brc$ and center-of-mass
 $\brc_{CM}$ coordinates and performed the integration over 
 $\brc_{CM}$ which eliminated the inverse volume prefactor
 in Eq.(\ref{eq:digamma}). By virtue of Eq.~(\ref{eq:enhartree}), the Hc energy functional 
$\ehc \den$ is thus replaced by the
classical electrostatic interactions of the corresponding charge
distributions and correlation contributions are neglected.

Evidently Eq.~(\ref{eq:enhartree}) can also be derived from a product
(mean-field) ansatz for the electron-nuclear part of the 
total wavefunction.
In fact, such a mean-field description of the electron-nuclear
interaction has been proposed in \cite{Thomas:69,ThomasJoy:70}
to study the protonic structure of molecules.

From the Hartree-energy functional (\ref{eq:enhartree}), the
corresponding potentials, defined in (\ref{eq:vnhcen}) and
(\ref{eq:vehcdef}), are readily calculated:
\begin{eqnarray}
  \label{eq:vnhen}
  V_{\rm H}^{en}(\brc) &=& \int d\br \; W_{en}(\brc,\br) \rho(\br) \\
  \label{eq:vehen}
  v_{\rm H}^{en}(\br) &=& \int d\brc \; W_{en}(\brc,\br) \Gamma (\brc) .
\end{eqnarray}
We note that, within the Hartree approximation, the nuclear response
potential (\ref{eq:vnrsp}) vanishes and the conditional potential
(\ref{eq:vncond}) is given by (\ref{eq:vnhen}).
With these Hartree potentials inserted in
expressions (\ref{eq:elec-dia}) and (\ref{eq:nuc-dia}),
the MCKS equations (\ref{eq:ndikseq}) and (\ref{eq:KSelec-dia}) are
solved self-consistently as described above.

\subsubsection{Results}

In the following, the application of the Hartree approximation to
the $H_2$ and $H_2^+$ molecules is discussed.
\begin{table}
 \begin{center}
    \vspace{3mm}
    \begin{tabular}{cccccc} 
      \hline \hline
      & 
      &\multicolumn{1}{c}{BO}  
      &\multicolumn{1}{c}{Hartree}
      &\multicolumn{1}{c}{OAO} 
      &\multicolumn{1}{c}{SAO} \\
      \hline 
      & $-E_0$                & 1.130 & 1.121 & 1.122 & 1.124 \\
      & $T_{S}$               &       & 1.069 & 1.063 & 1.049 \\
      & $E^e_{{\rm Hxc},e}$     &       & 0.627 & 0.625 & 0.623 \\
      & $-E^{en}_{\rm Hc}$    &       & 3.496 & 3.487 & 3.471 \\
      & $W_{nn}$              &       & 0.679 & 0.676 & 0.676 \\
      & $\langle R \rangle  $ & 1.49  & 1.48  & 1.49  & 1.50  \\
      & $\langle R^2 \rangle$ & 2.25  & 2.21  & 2.24  & 2.28  \\
      & $\omega$[cm$^{-1}$]   & 4137  & 7945  & 7047  & 4282  \\
      \hline \hline
    \end{tabular}
    \vspace{3mm}
    \caption{\label{tab:resh2} \it Summary of results for 
      the $H_2$ molecule obtained from self-consistent solutions
      of the MCKS scheme employing various approximations.
      For comparison, results from BO calculations are added.
      The electronic interaction is treated within the xcLDA. All numbers 
      (except $\omega$) in atomic units.}
  \end{center}
\end{table}
\begin{table}
  \begin{center}
    \vspace{6mm}
    \begin{tabular}{cccccc} 
      \hline \hline
      & 
      &\multicolumn{1}{c}{BO}  
      &\multicolumn{1}{c}{Hartree}
      &\multicolumn{1}{c}{OAO} 
      &\multicolumn{1}{c}{SAO} \\
      \hline 
      & $-E_0$                & 0.598 & 0.591 & 0.595 & 0.581 \\
      & $T_{S}$               &       & 0.591 & 0.583 & 0.574 \\
      & $-E^{en}_{\rm Hc}$    &       & 1.673 & 1.662 & 1.642 \\
      & $W_{nn}$              &       & 0.491 & 0.485 & 0.487 \\
      & $\langle R \rangle  $ & 2.07  & 2.05  & 2.08  & 2.08  \\
      & $\langle R^2 \rangle$ & 4.30  & 4.22  & 4.37  & 4.39  \\
      & $\omega$[cm$^{-1}$]   & 2297  & 5191  & 3248  & 2232  \\
      \hline \hline
    \end{tabular}
    \caption{\label{tab:resh2+} \it Summary of results for 
      the \htp molecule obtained from self-consistent solutions
      of the MCKS scheme employing various approximations.
      For comparison, results from BO calculations are added.
      All numbers (except $\omega$) in atomic units.}
    \vspace{3mm}
\end{center}
\end{table}

\begin{figure}
  \includegraphics[width=200pt]{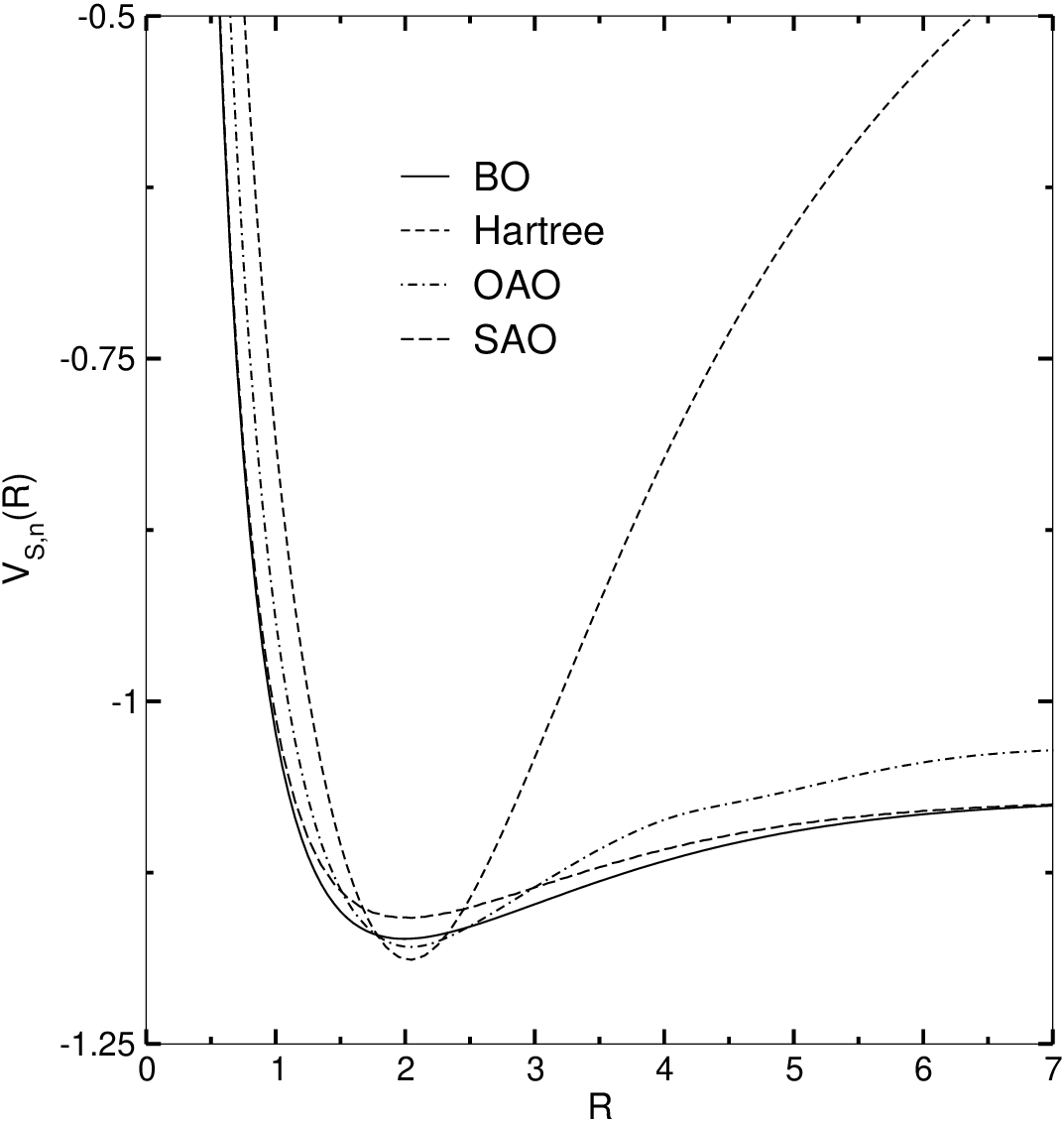}
  \caption{\label{fig:resh2+} \it Effective nuclear potential $V_{S,n}(R)$ 
    for the \htp molecule 
    obtained from self-consistent solutions
    of the MCKS scheme employing various approximations.
    For comparison, results from exact and BO calculations are added.
    In atomic units.}
\end{figure}

\begin{figure}
  \centering
  \includegraphics[width=200pt]{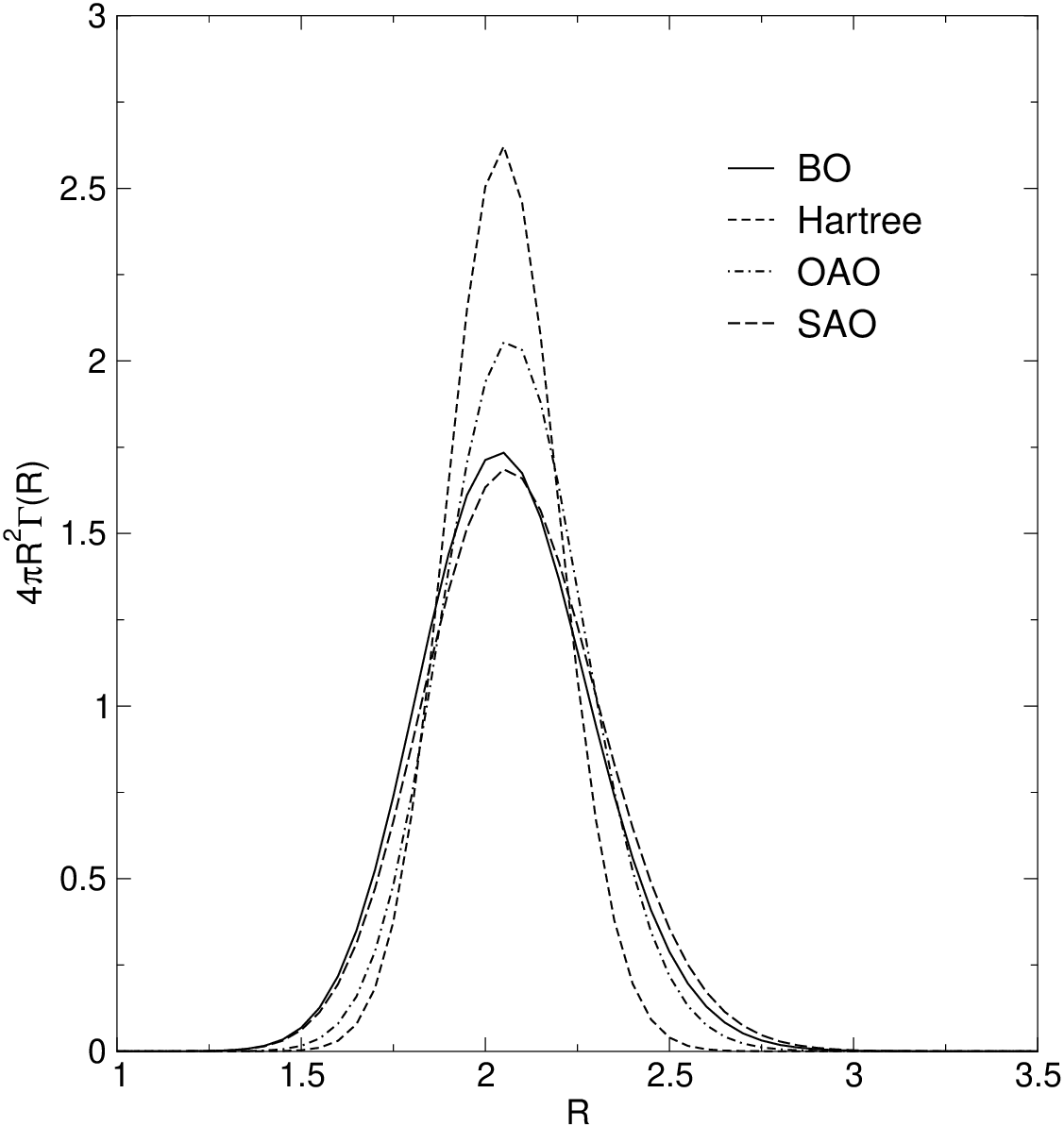}
 \caption{\label{fig:denh2+} \it Radial nuclear density $4 \pi R^2 \Gamma(R)$ 
    for the \htp molecule 
    obtained from self-consistent solutions
    of the MCKS scheme employing various approximations.
    For comparison, results from exact and BO calculations are added.
    In atomic units.}
\end{figure}

In tables~\ref{tab:resh2} and \ref{tab:resh2+} a selection of results is presented.
Since the BO approach provides an excellent approximation for the
system under consideration, we also added, for comparison, 
the results obtained from the BO calculation (employing the very 
same numerical procedure discribed above).
The ground-state results are found to be surprisingly good for both molecules.
Compared to the BO results listed in the first column of the tables,
the total ground-state energy and the geometry,
represented by the mean internuclear distance $\langle R \rangle$, are 
reproduced up to within an accuracy of about $1\%$ by the Hartree method.
However, turning towards the harmonic constants, we realize that the Hartree 
result is off by about a factor of two for both $H_2$ and $H_2^+$.
Since $\omega$ measures the curvature of the effective nuclear
potentials at the minimum, we may expect larger deviations in this
quantity.
\begin{figure}[ht]
  \includegraphics[width=200pt]{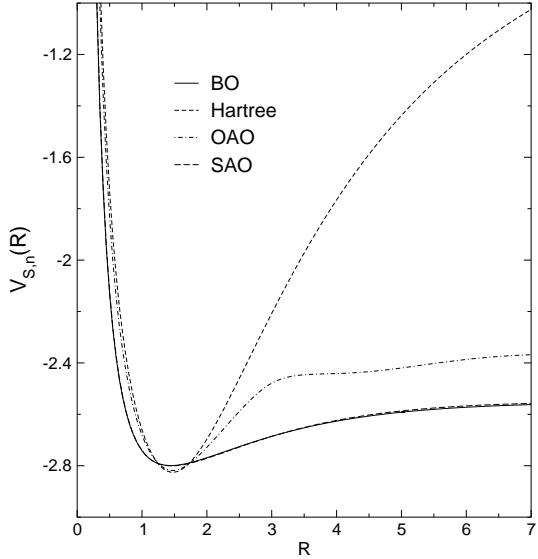}
  \caption{\label{fig:poth2} \it Effective nuclear potential $V_{S,n}(R)$
    for the $H_2$ molecule obtained from self-consistent solutions
    of the MCKS scheme employing various approximations.
    For comparison, results from a BO calculations is added.
    In atomic units.}
\end{figure}
\begin{figure}[ht]
  \includegraphics[width=200pt]{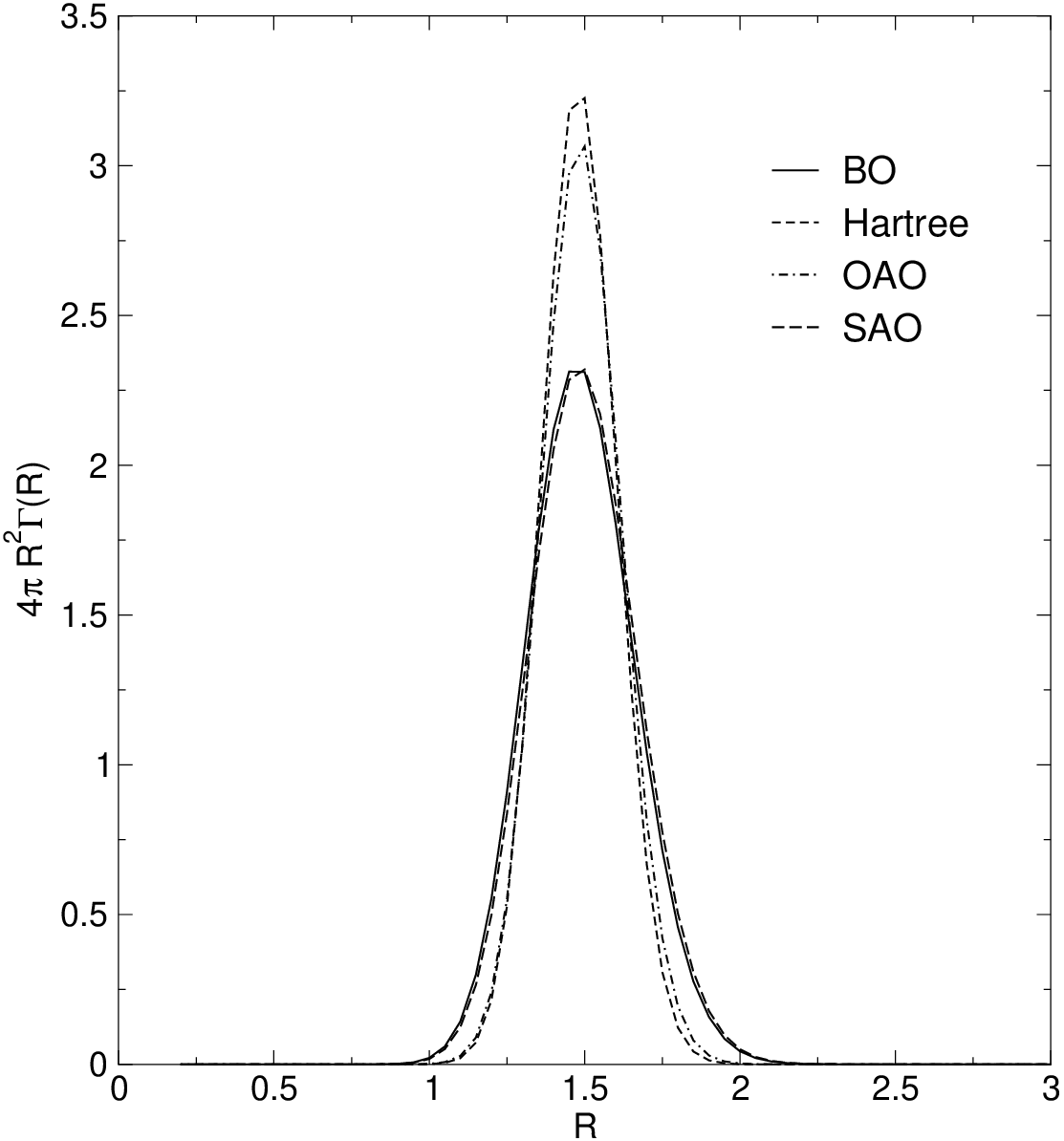}
  \caption{\label{fig:denh2} \it  Radial nuclear density $4 \pi R^2 \Gamma(R)$
    for the $H_2$ molecule obtained from self-consistent solutions
    of the MCKS scheme employing various approximations.
    For comparison, results from exact and BO calculations are added.
    In atomic units.}
\end{figure}
In fact, this is confirmed by Fig.~\ref{fig:poth2} and
Fig.~\ref{fig:resh2+}, where the effective 
nuclear MCKS potential is plotted for the $H_2$ and $H_2^+$ molecule.
Clearly, the Hartree potential is satisfactorily only in a small region
around the minimum of the potential - which, however, is sufficient for
good results for the total ground-state energy or geometry.
For larger $R$, the potential grows much to fast such that the depth of the 
potentials is largly overestimated.

In fact from the large $R$ behavior of Eq.(\ref{eq:vnhen}) we see 
that the asymptotic behavior of the effective potential in the Hartree
approximation is given by
\begin{equation}
V_{S,n} (\brc) \asneq (Z_1 Z_2 - N_e M_{\rm nuc}( \frac{Z_1}{M_2} +
\frac{Z_2}{M_1} )) \frac{1}{R}
\end{equation}
whereas the BO potential approaches a finite value in this limit.
As a consequence of the steep rise of the Hartree potential, the 
corresponding nuclear densities of both molecules, 
shown in Fig~\ref{fig:denh2+} and Fig~\ref{fig:denh2}
are much more localized than the BO ones and reflect the 
wrong shape of the effective nuclear potential.

\begin{figure}[t]
  \centering
  \includegraphics[width=170pt,angle=-90]{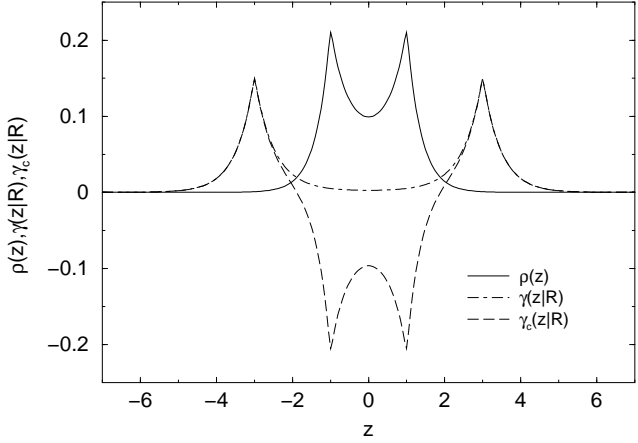}
  \caption{\label{fig:rhoc} \it  Typical behavior of the
    electronic density $\rho(z)$ and the conditional density
    $\gamma(z|R) \approx 6$~a.u.) for a diatomic molecule,
    plotted along the internuclear ($z$)~axis. In atomic units.}
\end{figure}

To understand the origin of these deviations in more detail, we reconsider
the expression for the Hxc energy functional, Eq.~(\ref{eq:ehc2}).
A comparison of this equation with the Hartree energy functional (\ref{eq:enhartree})
shows that the Hartree approximation for the electronic conditional
density is 
\begin{equation}
  \label{eq:rhoh}
  \gamma_{\rm H}(\br|\brc) = \rho(\br) \qquad \forall \, \brc,
\end{equation}
i.e., the conditional density is independent of $\brc$ in the
Hartree approach.
In the case that the nuclear density is a well-localized function, 
the approximation (\ref{eq:rhoh}) is justified for $\brc \approx \brc_{eq}$, 
(where $\brc_{eq}$ denotes the equilibrium separation) 
since many quantities of interest then only depend on internuclear
separations close to $\brc_{eq}$.
Thus the approximation (\ref{eq:rhoh}) leads to 
reasonable results, as reported above.
However, Eq.~(\ref{eq:rhoh}) fails for large $|\brc-\brc_{eq}|$.
This is illustrated in Fig.~\ref{fig:rhoc}, where we sketched a typical
behavior of the electronic density $\rho(\br)$ and the conditional density
$\gamma(\br|\brc)$ for $|\brc|\approx 6$~a.u.
As a consequence, the Hartree approximation cannot be expected to be
accurate for $R >| \brc_{eq}|$, explaining the deviations discussed in
Sec.~\ref{ssec:hartree}.

To summarize, the Hartree approximation provides a fair
estimate for ground-state properties of diatomic molecules,
such as the total ground-state energy or the equilibrium geometry.
However, the nuclear Hartree potential only reproduces the position of
the minimum but fails to correctly describe the shape of
the exact MCKS potential.
If one is interested in quantities depending more sensibly
on the shape of the nuclear potential, the Hartree method thus needs to be
improved.

\subsection{An Approach Based on Atomic Orbitals}
\label{ssec:lcao}

From the analysis of the preceding sections, we were lead to the conclusion 
that one needs to go beyond the Hartree approximation for the electron-nuclear
energy functional $\ehxc \den$.
Moreover, in view of the fact that the deviations between the Hartree and
the exact (BO) potentials are rather large, 
the Hartree potential is not necessarily a good starting point
for approximations -- in contrast to the situation for the electron-electron
interaction.
This is particularily true for properties which do not only depend
on internuclear distances close to $R_{eq}$.
For instance, to go beyond the Hartree approximation, it might be tempting 
(as usually done for electronic correlations in standard DFT) to define a "hole density"
\begin{equation}
\gamma_c (\br | \brc) := \gamma (\br | \brc) - \rho (\br)
\end{equation}
which measures the deviations of the density from the conditional density.
Inserting this definition into Eq.(\ref{eq:ehc2}) then leads to partitioning
of the Hartree-correlation functional into a Hartree and a correlation part.
However, 
as displayed in Fig.(\ref{fig:rhoc}) the hole function $\gamma_c (\br | \brc)$
is large for almost all $\brc$ (except $\brc \approx \brc_{\rm eq}$) and has to
account for large correlation corrections.
Therefore at the present point, it appears most promising to approximate 
the conditional density directly.
To that end, we first 
recall that $\gamma (\br|\brc)$ represents the probability
of finding an electron at $\br$, provided the nuclei are separated by 
$\brc$.
If we would go back to a BO description such a
quantity would be naturally described by the 
electronic density calculated within the BO approach as 
described in Eq.(\ref{eq:nclass1}) in which case we have
\begin{equation}
\gamma^{BO} (\br | \brc) = \sum_{j=1}^{N_e} | \xi^{BO}_{\brc,j} (\br )|^2
\label{eq:BOlimit}
\end{equation}
where $\xi^{BO}_{\brc,j}$ are the electronic single-particle
orbitals for internuclear distance $\brc$ as in standard density
functional theory.
In order to obtain Eq.(\ref{eq:BOlimit}) in the BO-limit 
we approximate the conditional density $\gamma$ in the spirit of a 
linear combination of atomic orbitals (LCAO) approach, i.e.,
\begin{equation}
\gamma (\br | \brc) \approx \sum_{j=1}^{N_e} \gamma_j (\br | \brc) 
\label{eq:conddenssum}
\end{equation}
where
\begin{equation}
  \label{eq:rhobo}
  \gamma_j (\br|\brc) = 
  \frac{1}{2 \nu_j(\brc)}
  \left| \phi^A_j (\br_A) + \phi^B_j (\br_B) \right|^2,
\end{equation}
where $\phi^{A/B}_j(\br)$ denotes an atomic-type orbital and 
the factor $\nu_j(\brc)$ is included to ensure normalization of the
orbital conditional density, i.e.
\begin{equation}
1 = \int d\br \; \gamma_j (\br | \brc) .
\label{eq:sumrule3}
\end{equation}
This ensures normalization of the electronic Kohn-Sham orbitals
in the BO limit.
Instead of following the standard LCAO approach where $\gamma(\br|\brc)$
is constructed from given atomic orbitals $\{\phi^{A/B}_j\}$,
we will determine suitably ``optimized'' atomic orbitals (OAO) from given
densities $\Gamma$ and $\rho$, i.e., the atomic orbitals are represented
as {\it functionals} of the densities: $\phi^{A/B}_j(\br) = \phi^{A/B}_j
[\Gamma,\rho](\br)$.
Inserting these orbitals in (\ref{eq:rhobo}) then leads to an 
approximation
of $\gamma[\Gamma,\rho]$ and therefore, by virtue of Eq.~(\ref{eq:ehc2}), 
to an approximation of $E^{en}_{\rm Hc}[\Gamma,\rho]$ as functionals
of the densities $\Gamma$ and $\rho$.

In order to find the OAO, we first note that, given {\it atomic orbitals}
$\phi^A_j(\br)$ and $\phi^B_j(\br)$, normalized bonding and antibonding
{\it molecular orbitals} for a fixed internuclear distance $R$ can
be obtained from
\begin{equation}
  \label{eq:moao}
  \p^{\pm}_{\brc,j}(\br) = \frac{1}{\sqrt{2 ( 1 \pm S_j(\brc) )}}
  \Big( \phi^A_j(\br_A) \pm \phi^B_j(\br_B) \Big),
\end{equation}
where $S_j(\brc) := \int d\br \; \phi^A_j(\br_A) \phi^B_j(\br_B)$
denotes the overlap integral and the atomic orbitals are assumed to
be real.
For the purpose of the section, we are concerned with the reverse
problem: Given molecular MCKS orbitals, how can we construct
corresponding atomic orbitals?

As the crucial idea, we identify the electronic MCKS orbitals
$\{ \p_j(\br) \}$, i.e., the solutions of the electronic MCKS equation
(\ref{eq:KSelec-dia}), with the bonding and antibonding orbitals
of Eq.~(\ref{eq:moao}) evaluated at the mean internuclear distance
$\langle R \rangle$:
\begin{eqnarray}
  \label{eq:identify}
  \p_j(\br) &\equiv& \p^+_{\langle R \rangle,j} (\br) \\
  \p_{\bar{j}}(\br) &\equiv& \p^-_{\langle R \rangle,j} (\br) ,
\end{eqnarray}
where $\p_{\bar{j}}$ denotes the antibonding counterpart of $\p_j$.
Using Eqs.~(\ref{eq:moao}) and (\ref{eq:identify}), we can solve
for the atomic orbitals, yielding
\begin{eqnarray}
   \phi^A_j[\Gamma,\{\p_j\}](\br) &=& \sqrt{\frac{1+S_j}{2}}
   \p_j \left(\br+\frac{M_2}{M_{\rm nuc}}\langle R \rangle {\bf e}_z 
\right) \nonumber \\
   &+& \sqrt{\frac{1-S_j}{2}}
   \p_{\bar{j}} \left(\br+\frac{M_2}{M_{\rm nuc}}\langle R \rangle
       {\bf e}_z \right)  \label{eq:ao1} \\
   \phi^B_j[\Gamma,\{\p_j\}](\br) &=& \sqrt{\frac{1+S_j}{2}}
   \p_j \left(\br-\frac{M_1}{M_{\rm nuc}}\langle R \rangle {\bf e}_z 
\right) \nonumber \\
   &-& \sqrt{\frac{1-S_j}{2}}
   \p_{\bar{j}} \left(\br-\frac{M_1}{M_{\rm nuc}}\langle R \rangle
       {\bf e}_z \right) .\label{eq:ao2} \\
\end{eqnarray}
The value of the overlap $S_j = S_j(\langle R \rangle )$ is not determined
by the above procedure and has to be supplied additionally.
A simple estimate is obtained from the overlap of unperturbed (hydrogenic)
orbitals.
Using this prescription, the atomic orbitals can be calculated from
Eqs.~(\ref{eq:ao1}) and (\ref{eq:ao2}).
We therefore determined the atomic orbitals as functionals of the
nuclear density and the electronic MCKS orbitals or, by virtue of the
MCHK theorem, as implicit functionals of the densities $\Gamma$ and $\rho$.

For the case of homonuclear diatomic molecules, the above equations
yield $\phi^A_j(\br) = \phi^B_j(-\br)$.
We note that the atomic orbitals are not required to have a definite
symmetry with respect to parity transformations.
As a matter of fact, we do not expect them to be symmetric; instead,
one may view them as orbitals which are centered on one nucleus and
polarized by the presence of the second nucleus.
Of course, for homonuclear molecules, the linear combinations
(\ref{eq:moao}) are properly symmetrized,
i.e., the molecular orbitals can be classified either as gerade or as
ungerade states.

Employing  Eqs.~(\ref{eq:ao1}) and (\ref{eq:ao2}), we readily obtain 
an approximation of the conditional density as an
(implicit) functional of the densities is obtained:
\begin{eqnarray}
  \label{eq:rholcao}
 \lefteqn{ \gamma^{\rm OAO}[\Gamma,\{\p_j\}](\br|\brc) }  \\
&=& \sum_{j=1}^{N_e}
  \frac{1}{2 \nu_j(\brc)} \left| \phi^A_j[\Gamma,\{\p_j\}] (\br_A)
    + \phi^B_j[\Gamma,\{\p_j\}] (\br_B) \right|^2. \nonumber
\end{eqnarray}
In the asymptotic $R \to \infty$ regime, Eq.~(\ref{eq:rholcao})
reduces to the correct asymptotic form, i.e., the conditional density
is given by the sum of two atomic densities
However, for the self-consistent ground-state solution of the
MCKS scheme employing Eq.~(\ref{eq:rholcao}), the atomic orbitals
are not the unperturbed orbitals representing the ground state
of the dissociated fragments,
but rather polarized orbitals which are optimized for the molecular
ground state.
Therefore, even for $R \to \infty$, the conditional density 
$\gamma^{\rm OAO}$ is not exact, although it should
improve on the Hartree behavior.

Employing Eq.~(\ref{eq:rholcao}), we obtain an expression for the
electron-nuclear interaction energy.
\begin{eqnarray}
  \label{eq:wen}
  \lefteqn{ W^{\rm OAO}_{en}[\Gamma,\rho]= \int d\brc \; \Gamma(\brc) } \\
&\times& 
  \int d\br \; W_{en}(\brc,\br) \; \gamma^{\rm OAO}
  [\Gamma,\{\p_j[\rho]\}](\br|\brc). \nonumber
\end{eqnarray}
In order to calculate the 
Hc energy functional from this expression we could 
perform the coupling-constant integration by using an approximation
for the $\lambda$-dependence of the conditional density
in Eq.(\ref{eq:ehc}).
Alternatively, we employ Eq.~(\ref{eq:ehcdef}), yielding
\begin{eqnarray}
  \label{eq:ehclcao}
\lefteqn{  E_{\rm Hc}^{\rm OAO}[\Gamma,\rho] = \int d\brc \; \Gamma(\brc) } \\
&\times& \left(
    \int d\br \; W_{en}(\brc,\br) \; \gamma^{\rm OAO}(\br|\brc)
    + F^e[\gamma^{\rm OAO}](\brc) - F^e[\rho] \right). \nonumber
\end{eqnarray}
where nonadiabatic terms have been neglected and $F^e$ is the
universal electronic functional defined in (\ref{eq:fedef}).
Eq.~(\ref{eq:ehclcao}) represents the central result of this section.
In order to use this approximation in a self-consistent MCKS calculation,
the effective potentials have to be calculated.
The nuclear conditional potential is readily evaluated, yielding
\begin{eqnarray}
  V_{\rm cond}^{\rm OAO}(\brc) &=& V_{\rm cond,W}^{\rm OAO}(\brc) +
  V_{\rm cond,T}^{\rm OAO}(\brc) \nonumber \\
&+&  V_{\rm cond,H}^{\rm OAO}(\brc) +
  V_{\rm cond,xc}^{\rm OAO}(\brc),
  \label{eq:vcondlcao}
\end{eqnarray}
where
\begin{eqnarray}
  \label{eq:vcondw}
  V_{\rm cond,W}^{\rm OAO}(\brc) &=&
  \int d\br \; W_{en}(\brc,\br) \; \gamma^{\rm OAO}(\br|\brc) \\
  \label{eq:vcondt}
  V_{\rm cond,T}^{\rm OAO}(\brc) &=& \frac{1}{8} \sum_j \int d\br \;
  \frac{| \nabla \gamma_j^{\rm OAO}(\br|\brc)|^2}
  {\gamma_j^{\rm OAO}(\br|\brc)} \nonumber \\
&-& T_{S,e}[\rho] \\
  \label{eq:vcondh}
  V_{\rm cond,H}^{\rm OAO}(\brc) &=& \frac{1}{2} \int \int  d\br d\br' \;
  \frac{\gamma^{\rm OAO}(\br|\brc) \gamma^{\rm OAO}(\br'|\brc)}
  {|\br-\br'|} \nonumber \\
&-& E^e_{\rm H}[\rho] \\
  \label{eq:vcondxc}
  V_{\rm cond,xc}^{\rm OAO}(\brc) &=&
  E^e_{\rm xc}[\gamma^{\rm OAO}](\brc) - E^e_{\rm xc}[\rho].
\end{eqnarray}
The first term on the right-hand side of Eq.~(\ref{eq:vcondlcao})
represents the functional derivative of $W^{\rm OAO}_{en}[\Gamma,\rho]$
with respect to the nuclear density.
The remaining terms are responsible for the $R$-dependence of the electronic
contributions to the Hxc energy functional.
The functional derivative of Eq.(\ref{eq:ehclcao})
is defined up to an arbitrary constant.
For this reason the terms (\ref{eq:vcondt})-(\ref{eq:vcondxc})
are defined such that they vanish 
when $\gamma (\br| \brc) \approx \rho (\br)$.
The nuclear response potential is neglected for the reasons explained
already above.
In order to calculate the electronic potential from (\ref{eq:ehclcao}),
one would have to resort to the optimized effective potential (OEP)
method \cite{SharpHorton:53,TalmanShadwick:76,GraboEtAl:99},
since the OAO energy
functional depends explicitly on the electronic MCKS orbitals $\{\p_j\}$
and therefore implicitly on the electronic density.
This, however, would lead to a rather complicated integral equation.
On the other hand, it was shown in Sec.~\ref{ssec:hartree} that the
electronic Hartree potential is sufficiently accurate for the
systems considered here.
Therefore, $v_{\rm H}^{en}(\br)$ will be used as an
approximation for the electronic Hc potential in the current
context, too.
Having derived the effective potentials, the MCKS scheme is solved
self-consistently.

\subsubsection{Results}

\begin{figure}[t]
  \centering
  \includegraphics[width=180pt,angle=-90]{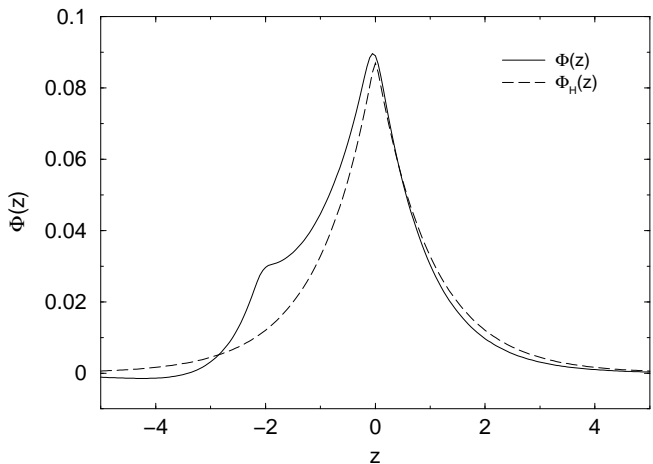}
  \caption{\label{fig:phiat} \it Atomic orbital obtained for the
    \htp molecule from a self-consistent solution of the OAO scheme
    explained in the text and compared to a hydrogenic (1s) orbital
    $\phi_H(z)$.
    Both curves are plotted along the electronic $z$ axis.
    In atomic units.}
\end{figure}
In the following, the results of the OAO appoach are presented for the
\htp and the $H_2$ molecules.
First of all, Fig.~\ref{fig:phiat} visualizes the optimized
atomic orbital as obtained from Eq.~(\ref{eq:ao1}) for the \htp molecular ion.
Compared to a hydrogenic $1s$ orbital, which is added to the plot
in dashed linestyle, we clearly see the anticipated influence of
the other nucleus:
At a distance of $R=\langle R \rangle \approx 2.2$~a.u.,
a second peak appears, leading to what we called a polarized orbital.
We may view this orbital as being optimized in the sense that
it provides the best ground-state solution when used in the
expression (\ref{eq:ehclcao}) for $E^{en}_{\rm Hc}$.
Indeed the results obtained for the $H_2$ and
$H_2^+$ molecules
molecule, which are again given in Tables~\ref{tab:resh2} and \ref{tab:resh2+},
consistently improve upon the Hartree data.
This remains true for the harmonic constant $\omega$, where the deviations
found in the Hartree scheme are somewhat reduced within the current approach.
Correspondingly, the  nuclear densities and potentials are slightly improved,
as seen in Figs.~\ref{fig:resh2+} - \ref{fig:denh2}.
However, the disagreement with the exact curves is still quite large.

\begin{figure}[t]
  \centering
 \includegraphics[width=200pt]{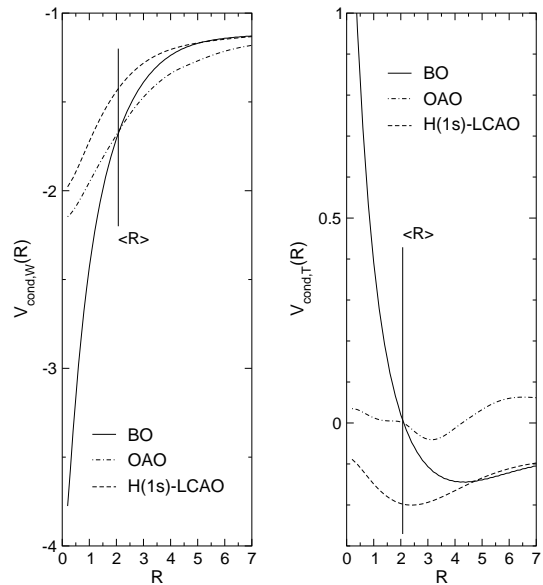}
  \caption{\label{fig:vw+t_h2+} \it Contributions to the nuclear
    conditional potential (\ref{eq:vcondlcao})
    as obtained from a self-consistent solution
    of the MCKS-OAO scheme for the \htp molecule. They are compared
    to the corresponding BO curves and to results provided by
    a simple LCAO employing hydrogenic (1s) orbitals. In addition, the
    mean (equilibrium) internuclear distance $\langle R \rangle$ is marked.
    In atomic units.}
\end{figure}
To further investigate this point, we have, following
Eq.~(\ref{eq:vcondlcao}), decomposed the conditional
potential into its different parts.
The results obtained for the \htp molecule are shown in
Fig.~\ref{fig:vw+t_h2+}, where $V_{\rm cond,W}(R)$, Eq.~(\ref{eq:vcondw}),
is plotted on the left-hand side and $V_{\rm cond,T}(R)$,
Eq.~(\ref{eq:vcondt}), is plotted on the right-hand side.
In addition to the curves obtained from the OAO and the BO approach,
which again serves as a reference, we added the
results calculated from a simple LCAO ansatz using hydrogenic ($1s$)
atomic orbitals in Eq.(\ref{eq:rhobo})
instead of the optimized atomic orbitals. The corresponding curve
is denoted by H($1s$)-LCAO in Fig.~\ref{fig:vw+t_h2+}.
We first observe that the results from the simple LCAO and the optimized
OAO scheme are very similar in shape.
Yet, the ones obtained from the optimized OAO are very close to
the exact (BO) numbers at the equilibrium internuclear
distance $\langle R \rangle$, as clearly visible in 
Fig.~\ref{fig:vw+t_h2+}.
This enables the OAO approach to predict ground-state
properties nicely, whereas the simple LCAO only leads to
qualitatively correct results.
However, since the optimized OAO potentials are basically shifted
(see Fig.\ref{fig:phiat}),
the $R \to \infty$ asymptotics, which -- by construction -- is
correctly described by the simple-OAO approach, is incorrect in the
present approach.
In the unified atom limit, on the other hand, both OAO schemes
produce large errors.
This is an inherent shortcoming of the OAO approach, which is not set up
to satisfy the $R\to 0$ limit.

In conclusion, the results of the method presented in this section
are clearly superior to the ones obtained from the Hartree approach.
Since the atomic orbitals are optimized for the molecular ground state,
the scheme works nicely for quantities depending mostly on the
equilibrium geometry but fails to substantially improve on the Hartree 
scheme for large internuclear distances.
This deficiency will be dealt with in the next section.

\subsection{Scaling the Atomic Orbitals}
\label{ssec:sao}

Owing to the successes of the OAO method to describe the bonding
region, an ansatz similar to (\ref{eq:rhobo}) will lay the foundation 
also for the work presented in this section.
However, in order to improve on the shortcomings of the OAO approach, we 
additionally concentrate on the task of setting up the scheme
such that the separated as well as the unified atom limit
are correctly reproduced.

The following investigations are again based on the quantity
which is considered to be the key quantity to approximate, namely the
electronic conditional density $\gamma(\br|\brc)$.
As above, we start out by decomposing the conditional density 
into orbital contributions as in Eq.(\ref{eq:conddenssum})
which are approximated by an LCAO-type ansatz
\begin{equation}
  \label{eq:rhosao1}
  \gamma_j(\br|\brc) \approx
  \frac{1}{2 \nu_j(\brc)}
  \left| \phi^A_j (\br_A) + \phi^B_j (\br_B) \right|^2,
\end{equation}
and $\br_{A/B} = \br \mp \frac{M_{2/1}}{M_{\rm nuc}} R {\bf e}_z$
as before.
The denominator $\nu_j(\brc)$ arises from the normalization 
constraint (\ref{eq:sumrule3}) and is given by
\begin{equation}
  \label{eq:nu}
  \nu_j(\brc) = \frac{1}{2} \int d\br \; \left| \phi^A_j (\br_A) 
    + \phi^B_j (\br_B) \right|^2.
\end{equation}
As a consequence, the first sumrule (\ref{eq:sumrule1}) for the
conditional density is
automatically satisfied for any choice of the atomic orbitals.

Up to this point, we just repeated the analysis of Sec.~\ref{ssec:lcao}.
Now, we have to specify the atomic orbitals $\{\phi^{A/B}_j\}$.
In the approach presented in the last section, they were determined
by optimizing the molecular ground-state configuration.
Since these orbitals were further used to describe the large-$R$
behavior of a diatomic molecule, the deviations reported on
above were found.

In order to improve on that, we consider the separated atom
($R \to \infty$) limit.
There, the system consists of two atoms $A$ and $B$, which do not
interact among each other.
The atoms can thus be described by electronic orbitals, denoted by
$\{\phi^{A/B}_{\infty,j}\}$, which yield the ground-state densities
$\rho_{A/B}$ of the fragments.
We assume that these orbitals are known, e.g., from an electronic
DFT calculation for the single atoms.
If the orbitals $\{\phi^{A/B}_{\infty,j}\}$ were inserted in 
Eq.~(\ref{eq:rhosao1}), we would obtain the simple LCAO scheme which
was used for comparison in the last section.
Obviously, the bonding effects are not satisfactorily described
within such a simple ansatz.
In view of the mutual influence of the atoms, 
we expect the orbitals to change when the atoms approach each other
as illustrated by Fig.~\ref{fig:phiat} in the last
section.

Here, we account for this change by introducing contracted orbitals
\begin{equation}
  \label{eq:phiatla}
  \phi^{A/B}_j(\br) \equiv  \phi^{A/B}_{\lambda,j}(\br)
  = \lambda_j^{3/2} \phi^{A/B}_{\infty,j}(\lambda_j \br).
\end{equation}
The idea to model bonding effects by such a scaling procedure
can be explained in terms of the virial theorem:
A decrease of the total energy due to chemical bonding leads to an
increase of the kinetic energy and thus to a contraction of the
orbitals \cite{Leeuwen:94}.
Evidently, the size of the contraction depends on the molecular
configuration and, in particular, on the internuclear distance.
It should therefore be determined self-consistently from the densities
which characterize the system, i.e.,
\begin{equation}
  \label{eq:lambda}
  \lambda_j = \lambda_j[\Gamma,\rho](\brc) .
\end{equation}
At this point, we already see some benefits of this approach.
If, for the equilibrium geometry, a scaling parameter $\lambda>1$ is used,
the description of molecular bonding is improved upon the simple 
LCAO approach.
On the other hand, employing $\lambda \to 1$ for $R \to \infty$, 
the large-$R$ asymptotics of the conditional density is correctly
reproduced.
In the following, we will describe a way to calculate the scaling
function $\lambda_j(\brc)$ within the MCKS scheme.
We start by employing the second sumrule (\ref{eq:sumrule2}).
To construct our functional it is assumed that this equation also holds true for its analogue 
formulated in terms of the orbital densities 
(as it does in the BO limit),
\begin{equation}
  \label{eq:rhosumorb2}
  \rho_j(\br) = \int d\brc \; \Gamma(\brc) \gamma_j(\br|\brc),
\end{equation}
where $\rho_j(\br) = |\p_j(\br)|^2$.
If Eq.~(\ref{eq:rhosumorb2}) is satisfied for all $j$, the sumrule
(\ref{eq:sumrule2}) is obeyed, too.
Employing Eqs.~(\ref{eq:rhosao1}) and (\ref{eq:phiatla}), 
Eq.~(\ref{eq:rhosumorb2}) is rewritten as 
\begin{eqnarray}
  \label{eq:rhosumorb2b}
\lefteqn{  \rho_j(\br) = \int d\brc \; \Gamma(\brc) \frac{\lambda_j^3(\brc)}{2 \nu_j(\brc)} } \\
 &\times&  \left| \phi^A_{\infty,j}\Big(\lambda_j(\brc) \br_A\Big) + 
  \phi^B_{\infty,j}\Big(\lambda_j(\brc) \br_B\Big) \right|^2 . \nonumber
\end{eqnarray}
Once the atomic orbitals $\{\phi^{A/B}_{\infty,j}\}$ are given,
the above integral equation determines $\lambda_j(\brc)$ as
an (implicit) functional of the nuclear density $\Gamma$ and of the
electronic MCKS orbital densities $\rho_j$.
However, a full solution of this integral equation is rather 
complicated and will not be attempted in the present approach.

Instead, a simplified scheme appears highly desirable.
To this end, we investigate the limits of the 
scaling function $\lambda_j(\brc)$.
As already noted above, we impose the condition that
\begin{equation}
  \label{eq:lambdartoinfty}
  \lambda_j(\brc) \asneq 1,
\end{equation}
which guarantees the correct $R \to \infty$ asymptotic behavior
of the conditional density $\gamma(\br|\brc)$.
Next, we consider the unified atom limit $R=0$.
In that case, the orbital conditional density reads
\begin{equation}
  \label{eq:rhor=0}
  \gamma_j(\br|\brc=0) = \frac{\lambda_j^3(0)}{2 \nu_j(0)}
  \left| \phi^A_{\infty,j}\Big(\lambda_j(0) \br \Big) + 
  \phi^B_{\infty,j}\Big(\lambda_j(0) \br\Big) \right|^2.
\end{equation}
This quantity should be equal to the electronic orbital density 
of the unified atom, $\rho_{A+B,j}(\br)$.
We therefore choose $\lambda_j(0)$ such that $\gamma(\br|0)$
most closely resembles $\rho_{A+B,j}(\br)$.
In other words, $\lambda_j(0)$ is obtained from the minimization
problem
\begin{equation}
  \label{eq:min1}
  \min_{\lambda_j(0)} \int d\br \; \Big| \gamma_j(\br|0) - \rho_{A+B,j}(\br)
    \Big|^2,
\end{equation}
where the unified-atom density is assumed to be known.

We illustrate this prescription for \htp molecule:
In the separated atom limit, the electron is represented by
a hydrogenic ($1s$) orbital
\begin{equation}
  \label{eq:h1s}
  \phi^A_{\infty}(\br) = \phi^A_{\infty}(\br) = \frac{1}{\sqrt{\pi}}
  \exp{(-r)},
\end{equation}
whereas the density of the unified -- in this example: $He^+$ -- atom
reads
\begin{equation}
  \rho_{A+A}(\br) \equiv \rho_{He^+}(\br) 
  = \frac{Z_{He^+}}{\pi} \exp{(-2 Z_{He^+} r)}.
\end{equation}
From Eq.~(\ref{eq:min1}), we immediately obtain
\begin{equation}
  \label{eq:lambdah2+}
  \lambda(0) = Z_{He^+}.
\end{equation}
Therefore, $\lambda(0)$ is given by the sum of the charges of the
two nuclei.
In a more complicated system, we expect the bare nuclear charge 
$Z_{A+B}$ to be replaced by an effective one.
Employing Eq.~(\ref{eq:lambdah2+}), it is easily seen that
the conditional density (\ref{eq:rhosao1})
reproduces the correct unified atom limit.

Furthermore, the small-$R$ behavior of the conditional density is
analyzed.
This can be done by expanding the electronic Hamiltonian $\hat{H}_e$
in powers of $R$, yielding
\begin{equation}
  \label{eq:heexpand}
  \hat{H}_e = \hat{H}_{A+B} + \frac{M_1Z_2 - M_2 Z_1}{M_{\rm nuc}}
  \sum_j \frac{z_j}{r_j^3} R + \mathcal{O}(R^2),
\end{equation}
where $\hat{H}_{A+B}$ denotes the Hamiltonian of the unified atom.
For homonuclear systems, to which we restrict ourselves in all
numerical calculations, the first-order correction in Eq.~(\ref{eq:heexpand})
vanishes.
From the fact that the electronic density corresponding to (\ref{eq:heexpand})
basically coincides with $\gamma(\br|\brc)$, we obtain the
small-$R$ ($R \to 0$) behavior:
\begin{equation}
  \label{eq:rhosmallR}
  \gamma(\br|\brc) = \rho_{A+A}(\br) +  \mathcal{O}(R^2).
\end{equation}

In view of the limits discussed above, we propose a simple
parameterization of the scaling function:
\begin{equation}
  \label{eq:lambdapara}
  \lambda_j(\brc) = 1 + \frac{\alpha_j}{1+\beta_j R ^{\gamma}}.
\end{equation}
Using such a form, the $R \to \infty$ limit (\ref{eq:lambdartoinfty})
is fulfilled.
The constant $\alpha_j$ follows from the unified atom limit,
Eq.~(\ref{eq:min1}): $\alpha_j = \lambda_j(0)-1$.
The exponent $\gamma$ is chosen such that
the model conditional density behaves -- for homonuclear molecules --
as (\ref{eq:rhosmallR}) for $R \to 0$, leading to $\gamma=2$.
We are therefore left with one still unknown coefficient, namely 
$\beta_j$.
To determine this constant, we resort to the integral equation
(\ref{eq:rhosumorb2}).
Employing additionally quasi-classical nuclei, $\Gamma(\brc)=\delta(\brc-\langle
\brc \rangle)$, we obtain
\begin{equation}
  \label{eq:sumsc}
  \rho_j(\br) = \gamma_j(\br|\langle \brc \rangle).
\end{equation}
The coefficient $\beta_j$ is then obtained self-consistently from fitting the 
model conditional density to the electronic MCKS orbitals densities:
\begin{equation}
  \label{eq:min2}
  \min_{\beta_j} \int d\br \; \Big| \rho_j(\br) - 
  \gamma_j(\br|\langle \brc \rangle) \Big|^2.
\end{equation}

Having calculated $\beta_j$, we put together all ingredients for the
construction of the model conditional density, which is denoted by
SAO (scaled atomic orbital) in the following, and
finally arrive at
\begin{eqnarray}
  \label{eq:rhosao}
\lefteqn{  \gamma^{\rm SAO}[\Gamma,\rho](\br|\brc) = \sum_j 
  \frac{\lambda_j^3(\brc)}{2 \nu_j(\brc)} } \\
&\times&  \left| \phi^A_{\infty,j}\Big(\lambda_j(\brc) \br_A\Big) + 
  \phi^B_{\infty,j}\Big(\lambda_j(\brc) \br_B\Big) \right|^2 , \nonumber 
\end{eqnarray}
with $\lambda_j(\br)$  given by Eq.~(\ref{eq:lambdapara}).
Summarizing the above prescription, the parameters in $\lambda_j$
are obtained from (i) the atomic orbitals corresponding to the
unified and the separate atom limit, which have to be provided as an input,
and (ii) self-consistently via Eq.~(\ref{eq:min2}) from the
MCKS (orbital) densities.
Thereby, the conditional density $\gamma^{\rm SAO}[\Gamma,\rho]$ is an 
(implicit) functional of the MCKS densities.
Moreover, the SAO conditional density now satisfies, by construction,
both normalization
sumrules (the second sumrule at least in a good approximation),
reproduces the correct asymptotic behavior and
hence meets all the requirements set up in the beginning of
the section.

Having obtained an approximation for the conditional density, we again have 
to face the problem of the coupling-constant integration
in Eq.~(\ref{eq:ehc2}).
One possibility to overcome this problem was discussed in Sec.~\ref{ssec:lcao},
where the Hc energy functional is expressed exclusively in terms of the 
conditional density at full coupling strength.
Employing this expression, Eq.~(\ref{eq:ehcdef}), we finally obtain
\begin{eqnarray}
  \label{eq:ehcsao}
  \lefteqn{ E_{\rm Hc}^{\rm SAO}[\Gamma,\rho] = \int d\brc \; \Gamma (\brc) } \\
&\times& \Big( 
    \int d\br \; W_{en}(\brc,\br) \; \gamma^{\rm SAO}(\br|\brc)
    + F^e[\gamma^{\rm SAO}](\brc) - F^e[\rho] \Big).\nonumber 
\end{eqnarray}
The corresponding nuclear conditional potential is given by
\begin{eqnarray}
  V_{\rm cond}^{\rm SAO}(\brc) &=& V_{\rm cond,W}^{\rm SAO}(\brc) +
  V_{\rm cond,T}^{\rm SAO}(\brc) \nonumber \\
&+& V_{\rm cond,H}^{\rm SAO}(\brc) +
  V_{\rm cond,xc}^{\rm SAO}(\brc),  \label{eq:vcondsao}
\end{eqnarray}
with
\begin{eqnarray}
  \label{eq:vcondwsao}
  V_{\rm cond,W}^{\rm SAO}(\brc) &=& 
  \int d\br \; W_{en}(\brc,\br) \; \gamma^{\rm SAO}(\br|\brc) \\
  \label{eq:vcondtsao}
  V_{\rm cond,T}^{\rm SAO}(\brc) &=& \frac{1}{8} \sum_j \int d\br \;
  \frac{| \nabla \gamma_j^{\rm SAO}(\br|\brc)|^2}
  {\gamma_j^{\rm SAO}(\br|\brc)} \nonumber \\
  &-& T_{S,e}[\rho] \\
  \label{eq:vcondhsao}
  V_{\rm cond,H}^{\rm SAO}(\brc) &=& \frac{1}{2} \int \int  d\br d\br' \;
  \frac{\gamma^{\rm SAO}(\br|\brc) \gamma^{\rm SAO}(\br'|\brc)}
  {|\br-\br'|} \nonumber \\
  &-& E^e_{\rm H}[\rho] \\
  \label{eq:vcondxcsao}
  V_{\rm cond,xc}^{\rm SAO}(\brc) &=& 
  E^e_{\rm xc}[\gamma^{\rm SAO}](\brc) - E^e_{\rm xc}[\rho].
\end{eqnarray}
As above, Eq.~(\ref{eq:vcondsao}) can be used in the MCKS scheme.

As an interesting aside, we add an alternative approach to
calculate the nuclear conditional potential directly from
$\gamma^{\lambda=1}(\br|\brc)$.
The idea rests on the observation that 
$V_{\rm cond}(\brc)$ is practically identical to the lowest energy
BO-PES, if non-BO effects are negligible.
Employing the Hellmann-Feynman (electrostatic) theorem 
\cite{Feynman:39}, we obtain
\begin{eqnarray}
  \frac{\partial V_{\rm cond}(\brc)}{\partial \brc} &\equiv& 
  \frac{\partial \epsilon^{\rm BO}(\brc)}{\partial \brc} = 
    \langle  \frac{\partial \hat{W}_{\rm en}}{\partial \brc} \rangle \nonumber \\
  &=& \int d\br\; \frac{\partial W_{\rm en}(\brc,\br)}{\partial \brc}
    \gamma(\br|\brc) .   \label{eq:hft}
\end{eqnarray}
Evidently the slope of the nuclear conditional potential is solely determined
by the conditional density at full coupling strength and could therefore be
evaluated using Eq.~(\ref{eq:rhosao}).
Moreover, compared to Eq.~(\ref{eq:vcondsao}), the expression (\ref{eq:hft}) 
seems to be more efficient from a numerical point of view.
However, the first approach leading to
Eq.~(\ref{eq:vcondsao}) proved to be more accurate and will therefore be used 
in the calculations presented below.

To summarize, the nuclear conditional potential is calculated from
Eq.~(\ref{eq:vcondsao}) by using the model SAO conditional density
(\ref{eq:rhosao}).
As for the OAO approach, we furthermore neglect the nuclear response
potential, approximate the electronic potential by the Hartree ansatz, and
solve the MCKS scheme self-consistently.

\subsubsection{Results}

\begin{figure}[b!]
  \centering
  \includegraphics[width=180pt,angle=-90]{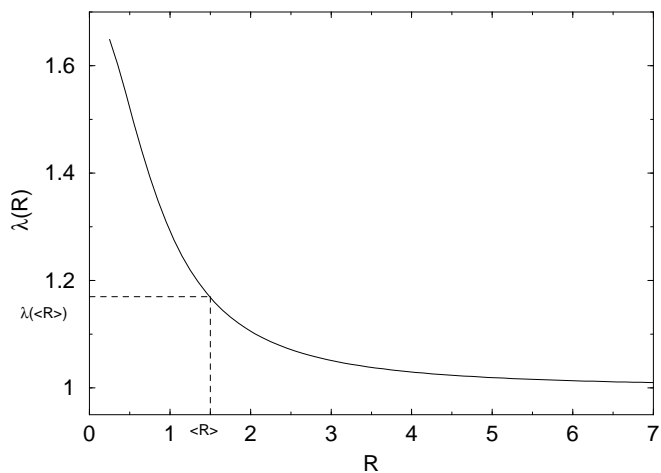}
  \caption{\label{fig:lambda} \it Scaling function $\lambda(R)$,
    Eq.~(\ref{eq:lambdapara}), obtained from a self-consistent solution
    of the MCKS scheme for the $H_2$ molecule.
    In atomic units.}
\end{figure}
First of all, we consider the scaling function $\lambda(R)$ obtained
from a self-consistent MCKS calculation for the $H_2$ molecule, 
which is plotted in Fig.~\ref{fig:lambda}.
By construction, $\lambda(R)$ tends to one for large $R$
such that the dissociation limit is correctly reproduced.
For small $R$, on the other hand, we find that $\lambda(0) \approx 1.7$.
As expected above, this number is similar to the effective charge
one obtains for the $He$ atom within a simple Hartree-Fock treatment
employing hydrogenic orbitals.
At the equilibrium distance $\langle R \rangle$, 
the scaling function acquires a value of about
$1.17$, leading to a contraction of the orbital by this factor.
As a consequence, the bonding energy is lowered compared to a simple
LCAO ansatz.
Indeed, the ground-state energy of the $H_2$ molecule obtained from the 
SAO approximation is close to the exact one, as is seen from the last
column of Tab.~\ref{tab:resh2}.
The energy improved on the Hartree and on the LCAO data, reducing the
error to about $0.5\%$.
We also observe that the $R$-expectation values are slightly 
overestimated, which can be viewed as a left-over from the simple 
LCAO method, which generally tends to overestimate the bonding distances.
The remaining deviations can be attributed to changes in the orbital
like, e.g., the appearance of a second peak as seen in Fig~\ref{fig:phiat},
which cannot be accounted for by the simple scaling procedure used in
the SAO scheme presented here.
This effect seems to be more pronounced for the \htp molecule.
From Tab.~\ref{tab:resh2+} we find that
especially the ground-state energy is somewhat worse than the results
obtained from the other approximations.
However, we find a remarkable improvement in the harmonic constant.
For both the \htp as well as the $H_2$ molecule, the relative error
in $\omega$ is lowered by more than an order of magnitude to
about $3\%$.
\begin{figure}[t]
  \centering
  \includegraphics[width=200pt]{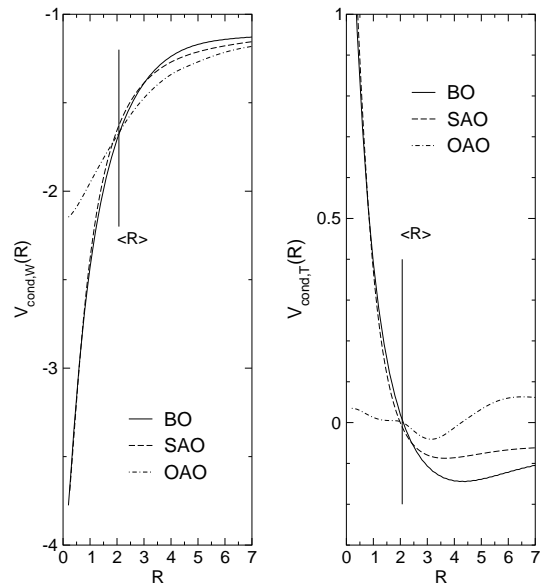}
  \caption{\label{fig:vw+tsao_h2+} \it Comparison of  
    different contributions to the nuclear potential,
    Eqs.~(\ref{eq:vcondwsao}) and (\ref{eq:vcondtsao}), as obtained
    for the \htp molecule from the BO, the SAO, and the OAO method. 
    In atomic units.}
\end{figure}
Correspondingly, the nuclear densities and potentials are expected to be
closer to the exact results, too.
From Figs.~\ref{fig:resh2+} and \ref{fig:poth2},
we indeed find that the nuclear potentials obtained from the SAO approach 
are almost indistinguishable from the BO-PES in the 
asymptotic ($R \to 0$ and $R \to \infty$) regime.
Of course, this is hardly surprising, since the correct
asymptotic behavior was imposed in the construction of the conditional
density.
However, this consequently leads to a much improved shape of the nuclear
potential, which is obvious from a comparison of the SAO curves to the
Hartree or LCAO data and is also reflected in the better 
harmonic constant reported above.
Moreover, considering the $H_2$ molecule, the SAO
nuclear potential coincides with the BO-PES not only asymptotically, but
in the whole $R$ range.
This then leads to a nuclear density, plotted
in Fig.~\ref{fig:denh2}, which nicely agrees with the BO one.
For the \htp molecule, the agreement at intermediate internuclear
distances is not as good.
From Fig.~\ref{fig:denh2+}, we find that the nuclear density 
is slightly shifted to larger $R$.
Additionally, the minimum of
the SAO nuclear potential is too high, leading to the deviations in the
energy already mentioned in the discussion of Tab.~\ref{tab:resh2+}.
Still, the SAO approach provides the best overall description also
for the \htp molecule.
In particular, the nuclear density obtained from the SAO approach
is closest to the exact one.
Furthermore, as seen from Fig~\ref{fig:vw+tsao_h2+} where the 
different contributions to the nuclear conditional potential,
Eqs.~(\ref{eq:vcondwsao}) and (\ref{eq:vcondtsao}), are shown,
the SAO curves are in satisfactory agreement
with the BO results in the whole range of internuclear distances
and thus improve on the OAO method, which only
reproduces the correct values around the equilibrium distance.
We therefore observe the effect of incorporating the
unified as well as separated atom limit into the construction
of the Hc energy functional.
At this point, we also emphasize the importance to account
for the additional contributions which arise from  
the coupling-constant integration.
As is seen from the right-hand side of Fig~\ref{fig:vw+tsao_h2+}, 
the $R$-dependence of these terms is significant, and it would not
be a good approximation to replace $E_{\rm Hc}^{en}$ by $W_{en}$.
In conclusion we find that the SAO approximation to the electron-nuclear
correlation functional gives a very good overall description
of the BO potential. 

\section{Conclusions}

For a unified quantum mechanical treatment of nuclear and
electronic degrees of freedom we extended the traditional
density functional method to multicomponent systems.
We first discussed the choice of appropriate densities serving
as fundamental variables of the theory. It was shown that the
usual definition of single-particle densities in terms of
inertial coordinates is not well suited for the purpose of this
work because such densities, as a consequence of Galilean
invariance, are constant for all isolated systems and therefore
not characteristic for their internal properties.
A suitable set of densities was obtained by defining the
electronic density 
with respect to a coordinate system attached to the
nuclear framework whereas the nuclear degrees of freedom
were described by the diagonal of the nuclear $N_n$-body
density matrix. For these fundamental variables the
Hohenberg-Kohn theorem and the Kohn-Sham equations were
derived and the corresponding density functionals
were analyzed in detail. The main new ingredient of the 
multicomponent theory is the electron-nuclear
correlation functional. For this functional several
approximations were derived and tested on the $H_2$ and
$H_2^+$ molecules. It was found that the simplest
Hartree approximation fails to give a good description of the
bonding curve of these molecules. Considerable improvement was
obtained using an approximation based on optimized atomic orbitals.
This method still had some deficiencies in the large and
small bond distance limits. These deficiencies were finally removed
using an approximation based on scaled atomic orbitals.
Based on this first experience with MCDFT we can say that 
it presents a promising new approach to study electron-nuclear
correlation phenomena beyond the BO approximation.
A promising new field of applications seems the first
principle treatment of electron-phonon interactions
within a linear response language using a time-dependent
extension of the present theory. Another field of future applications will be 
the study of combined ionization and dissociation dynamics of
molecules in strong laser fields. For this case some approximations
of similar spirit as discussed in this paper have already been applied
succesfully~\cite{Kreibichetal:CP04,TDDFTbook}.

\section{Acknowledgements}
RvL acknowledges support of 'Stichting voor
Fundamenteel Onderzoek der Materie (FOM)'.



\end{document}